\documentclass[11pt,a4paper,fleqn]{article}

\usepackage{graphicx}
\usepackage{amsmath, amssymb,wasysym} 
\usepackage[T1]{fontenc}
\usepackage{helvet}
\usepackage[textwidth=175mm,textheight=240mm]{geometry}
\usepackage[width=0.9\textwidth]{caption}
\usepackage{fleqn}
\usepackage{color}
\usepackage{setspace}

\definecolor{mygray}{gray}{.75}
\definecolor{mygreen}{rgb}{0.0,0.48,0.16}

\usepackage[numbers,sort&compress]{natbib}

\usepackage{dcolumn}
\usepackage{bm}

\usepackage{amsmath}
\usepackage{amssymb}
\usepackage{amsfonts} 
\usepackage{graphics}  
\usepackage{graphicx} 
\usepackage{epsfig}    
\usepackage{rotating}  
\usepackage{color}
\usepackage{rotating}
\usepackage{multirow}
\usepackage{csquotes}
\usepackage[nooneline, tight,raggedright,FIGTOPCAP]{subfigure}
\usepackage[colorlinks=false]{hyperref}

\usepackage[open, openlevel=0]{bookmark}
\usepackage{enumitem}
\definecolor{darkgreen}{rgb}{0,0.5,0} 
\definecolor{violet}{rgb}{0.5,0,0.5}
\definecolor{orange}{rgb}{0.2,0.5,0.5}
\definecolor{darkgray}{rgb}{0.5,0.5,0.5}

\newcommand{\bequ}{\begin{equation}}
\newcommand{\eequ}{\end{equation}}
\newcommand{\bequa}{\begin{eqnarray}}
\newcommand{\eequa}{\end{eqnarray}}
\newcommand{\bse}{\begin{subequations}}
\newcommand{\ese}{\end{subequations}}
\newcommand{\tn}[1]{\textnormal{#1}}

\newcommand{\m}{microtubule}
\newcommand{\ms}{microtubules}
\newcommand{\mms}{molecular motors}

\usepackage{mathtools}

\usepackage{wrapfig}

\begin{document}


\title{\Large{\bf{Molecular Mechanisms for Microtubule Length Regulation by Kinesin-8 and XMAP215 Proteins}}}

\author{\normalsize{Louis~Reese$^{1,2}$} \and \normalsize{Anna~Melbinger$^{1,3}$} \and \normalsize{Erwin~Frey$^{1,2}$}\thanks{Corresponding author. Email:~frey@lmu.de} \\ \\
\normalsize{$^1$ Arnold Sommerfeld Center for Theoretical Physics and  Center for NanoScience,}\\ \normalsize{Department of Physics, Ludwig-Maximilians-Universit\"{a}t M\"{u}nchen,} \\ \normalsize{Theresienstra\ss e 37, D-80333 Munich, Germany} \\ \normalsize{$^2$ Nanosystems Initiative Munich (NIM), Ludwig-Maximilians-Universit\"{a}t M\"{u}nchen,} \\ \normalsize{Schellingstra{\ss}e 4, D-80333 Munich, Germany}\\ \normalsize{$^3$ Department of Physics, University of California,  San Diego, CA 92093, USA} }

\date{\normalsize{May 22, 2014}}

\maketitle
\vspace{-.5cm}

\begin{abstract}
The cytoskeleton is regulated by a plethora of enzymes that influence the stability and dynamics of cytoskeletal filaments. How microtubules are controlled is of particular importance for mitosis, during which dynamic microtubules are responsible for proper segregation of chromosomes. Molecular motors of the kinesin-8 protein family have been shown to depolymerise microtubules in a length-dependent manner, and recent experimental and theoretical evidence suggest a possible role for kinesin-8 in the dynamic regulation of microtubules. However, so far the detailed molecular mechanisms how these molecular motors interact with the growing microtubule tip remain elusive. Here we show that two distinct scenarios for the interactions of kinesin-8 with the microtubule tip lead to qualitatively different microtubule dynamics including accurate length control as well as intermittent dynamics. We give a comprehensive analysis of the regimes where length-regulation is possible and characterise how the stationary length depends on the biochemical rates and the bulk concentrations of the various proteins. For a neutral scenario, where microtubules grow irrespective of whether the microtubule tip is occupied by a molecular motor, length regulation is possible only for a narrow range of biochemical rates, and, in particular, limited to small polymerisation rates. In contrast, for an inhibition scenario, where the presence of a motor at the microtubule tip inhibits microtubule growth, the regime where length regulation is possible is extremely broad and includes high growth rates. These results also apply to situations where a polymerising enzyme like XMAP215 and kinesin-8 mutually exclude each other from the microtubule tip. Moreover, we characterise the differences in the stochastic length dynamics between the two scenarios. While for the neutral scenario length is tightly controlled, length dynamics is intermittent for the inhibition scenario and exhibits extended periods of microtubule growth and shrinkage reminiscent of microtubule dynamic instability. On a broader perspective, the set of models established in this work quite generally suggests that mutual exclusion of molecules at the ends of cytoskeletal filaments is an important factor for filament dynamics and regulation.
\end{abstract}

\section{Introduction}
\vspace{-.3cm}
Microtubules are essential constituents of the cytoskeleton of eukaryotic cells. They provide mechanical support, and are involved in a wide range of cellular functional modules. For instance, they serve cells to build cilia and flagellae which are slender extensions of the cell used for migration and sensory tasks. In addition, \ms~are important during cell division, where they build the mitotic spindle, and separate chromosomes. To facilitate this variety of tasks, there have to be mechanisms that  control for the dynamics of \ms~\cite{Desai1997}. Such capabilities are crucial for the \m~cytoskeleton in order to accomplish such diverse tasks as cell division~\cite{Goshima2010} and migration~\cite{deForges2012}, and further, to determine cell size and shape~\cite{Tischer2009}, and to position the nucleus in the center of the cell~\cite{Laan2012,Pavin2012}. Molecular motors and \m~associated proteins seem to play a crucial role in this regulation process~\cite{Howard2007,Wordeman2005}: Motors move along the microtubule and interact specifically with the filament at its end. Other associated proteins that influence the dynamics of \ms~also bind directly to the \m~tip~\cite{Akhmanova2008}. Biochemical reconstitution experiments with \ms~and \m~associated molecules have lead to considerable insight in recent years~\cite{Subramanian2012}. In the following we highlight two specific proteins, that are important for the length dynamics of~\ms.

Kip3p is a \m~depolymerising molecular motor~\cite{Varga2006,Gupta2006} of the kinesin-8 protein family~\cite{Walczak2013}. It binds strongly to the \m~lattice, and, therefore, exhibits a long run-length~\cite{Leduc2012}. At the \m~tip the effect of strong motor binding to the terminal tubulin heterodimer induces depolymerisation~\cite{Cooper2010}. Interestingly, kinesin-8 shows a length-dependent depolymerisation activity mediated by the accumulations of motors along the microtubule, as shown in experiments~\cite{Varga2006,Varga2009} and recent theoretical work~\cite{Reese2011}. In vitro experiments have unveiled many molecular details of kinesin-8: The tail of the motor has been shown to be responsible for long residence times on the \m~lattice and it influences \m~dynamics~\cite{Stumpff2011,Su2011,Mayr2011} and spindle size~\cite{Weaver2011}; for a brief review of these findings see Ref.~\cite{Su2012}. In the mathematical analysis we will concentrate on the depolymerising activity of the motor and treat those molecular details effectively in terms of rate constants for movement, depolymerisation activity, and attachment/detachment kinetics of the motors.

XMAP215 is a \m~associated protein~\cite{Gard1987}, that has been shown to significantly amplify the growth rate of \ms~\cite{Vasquez1994}, and to influence the dynamic properties of~\ms~in the cytosol~\cite{Tournebize2000,Kinoshita2002}. Recent in vitro experiments investigated the interaction of XMAP215 with single \ms~\cite{Brouhard2008} and the interplay with other end-binding proteins, which act as cofactors~\cite{Li2012,Zanic2013}. Specifically, it was found that XMAP215 is a polymerising enzyme to the \m~plus-end; a single XMAP215 is able to polymerise several rounds of tubulin heterodimers to the \m~\cite{Brouhard2008}. Similar properties have also been observed for other end-binding proteins, see \emph{e.g.}\ Ref.~\cite{Komarova2009}.

Combining the observations described above, suggests that kinesin-8 and XMAP215 may constitute a minimal functional unit able to regulate \m~dynamics~\cite{Tournebize2000,Kinoshita2001} and antagonistically influence \m~length. This view is supported by recent experiments on cilia~\cite{Niwa2012} showing that the molecular motor Kif19a, which belongs to the kinesin-8 protein family, regulates the cilia length in a concentration dependent manner: High motor concentrations lead to short cilia, whereas low motor concentrations ensue long cilia. On a molecular scale, the ability to regulate length is traced back to the observed length-dependent depolymerisation speed~\cite{Howard2007}. In more detail, longer \ms~are observed to depolymerise faster than shorter ones. This has been explained as follows~\cite{Varga2009,Reese2011}: Molecular motors in the cytosol attach to the \m~and subsequently move towards the \m~tip. The unidirectional movement of motors towards the \m~tip leads to an  accumulation of motors and the motor density increases from the minus- to the plus-end of the~\m, which finally results in an \emph{antenna-like} steady state profile of molecular motors. 
Therefore, there are more motors present at the tip for longer \ms~than for shorter ones, which in turn leads to the observed length-dependence in the depolymerisation speed. 
In combination with \m~polymerisation, which is either spontaneous or catalysed by XMAP215, this is a promising starting point to achieve \m~regulation~\cite{Howard2007,Melbinger2012}.
 
In this work, we elaborate on two possible molecular mechanisms how \mms~could interact with the \m~tip. We specifically distinguish two scenarios, one where \mms~prevent the addition of tubulin heterodimers at the \m~tip (inhibition scenario), and another neutral scenario where \m~growth is possible irrespective of whether the \m~tip is occupied by motors or not. These differences in the interaction of motors with the \m~tip give rise to a rich dynamics of \m~length ranging from accurate length-control to intermittent dynamics. 

This article is organised as follows. 
In section~\ref{sec:model} we introduce a model for the dynamics of molecular motors on a \m. Further we define different possible molecular scenarios for how kinesin-8 interacts with the \m~tip during the depolymerisation process, including the case when XMAP215 acts as a polymerase. 
In section \ref{sec:dynamics} we present our main results: We begin with an outline of the theoretical framework, and then employ it to study MT length dynamics. Our analytical calculations are complemented by stochastic simulations. Taken together, this allows us to identify the parameter regimes where length regulation is possible, and to provide a comprehensive analysis on how the ensuing stationary length depends on biochemical rates and protein concentrations. Moreover, we investigate the role of stochastic effects in length regulation, and discuss why there are dramatic differences between the considered scenarios.
Finally, we conclude in section~\ref{sec:discussion} by discussing our results in terms of their possible biological relevance and their importance for driven diffusive lattice gases.

\begin{figure*}[th]
\centering
\includegraphics[scale=1]{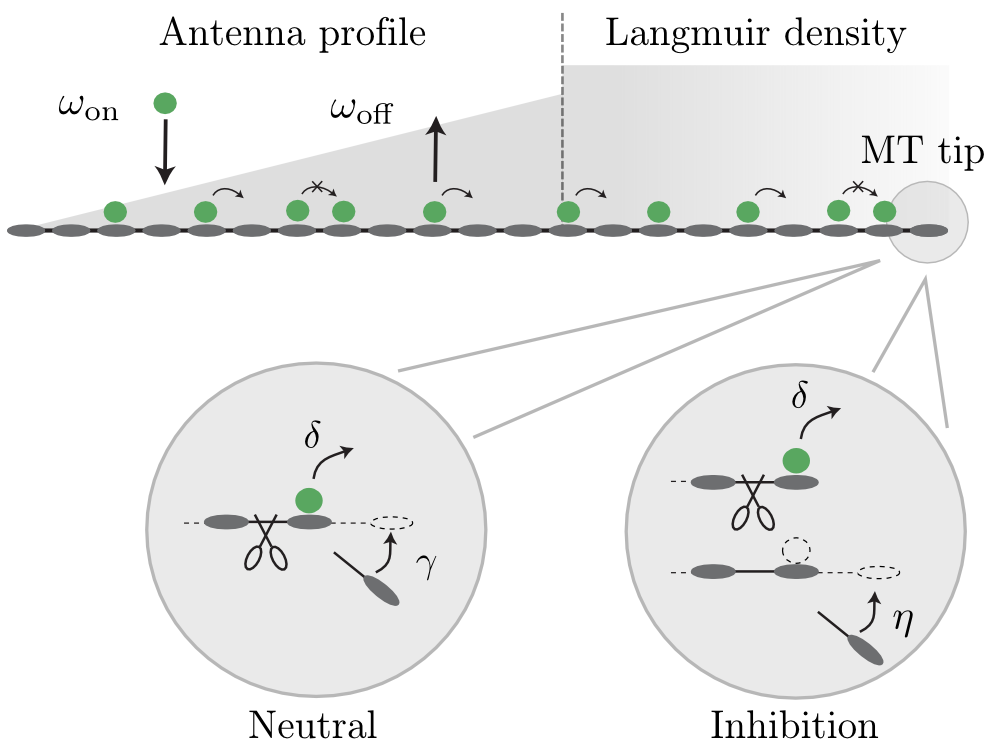}
\caption{
Illustrations of motors on a microtubule (MT) and different regulation scenarios at the MT plus-end. Starting from an empty MT lattice motors start to accumulate on the MT lattice by Langmuir kinetics (with rates $\omega_\tn{on}$ and $\omega_\tn{off}$) and subsequent transport with rate $\nu$ to the plus-end. The combined effect of Langmuir kinetics and steric exclusion between the motors leads to a antenna-like profile which saturates at the Langmuir density $\rho_\tn{La}$. At the MT tip, kinesin-8 depolymerises the MT lattice and blocks MT growth the \emph{inhibition scenario} while it does not affect MT growth in the \emph{neutral scenario}. \label{fig:cartoon}}
\end{figure*}

\section{Model definition}
\label{sec:model}\vspace{-.3cm}
We consider a one-dimensional lattice gas model of finite length $L$~\cite{Krapivsky2010,Chou2011} as illustrated in Fig.~\ref{fig:cartoon}: Motor proteins (kinesin-8), present at a constant bulk concentration $c$, are assumed to randomly attach to and detach from the microtubule (MT) lattice with rates $\omega_\text{on}$ and $\omega_\text{off}$, respectively, defining the binding constant $K=c \, \omega_\text{on}/\omega_\text{off}$. 
Once bound, these motors move towards the plus-end at a constant hopping rate $\nu$; we fix the time scale by setting $\nu = 1$ (corresponding to approximately $6.35$ steps/sec in the case of kinesin-8~\cite{Varga2009}).
As these motors hinder each other sterically, individual binding sites on MTs can at most be occupied once. This lattice gas model is known as the \emph{totally asymmetric simple exclusion process} (TASEP) with \emph{Langmuir kinetics} (LK)~\cite{Lipowsky2001,Parmeggiani2003,Parmeggiani2004}.

The right boundary is considered to be dynamic: When a kinesin-8 motor arrives at the MT plus-end, which is the boundary in our model, it acts as a depolymerase, \emph{i.e.}\ it removes the last MT subunit at rate $\delta$~\cite{Reese2011}. In addition, the MT is assumed to polymerise through the attachment of single tubulin heterodimers. Unfortunately, there is insufficient experimental information on the detailed molecular cycle for MT growth in the presence of kinesin-8 motors. We hypothesise the following different but equally plausible mechanisms for MT growth:
\begin{enumerate}[label=(\roman{enumi})]
 \renewcommand{\labelenumi}{(\roman{enumi})}
  \renewcommand{\theenumi}{(\roman{enumi})}
\item The MT only grows at rate~$\eta$ if the last site at the plus-end is \emph{not} occupied by a kinesin-8 motor. Because kinesin-8 inhibits MT growth we call this the \emph{inhibition scenario}; cf. Fig.~\ref{fig:cartoon}. 
\label{eta}
\item The MT grows at rate~$\gamma$ independently of whether the tip is occupied or not. This \emph{neutral scenario} has been considered previously in Ref.~\cite{Melbinger2012}; cf. Fig.~\ref{fig:cartoon}. \label{gamma}
\item MT polymerisation is facilitated by a second protein species, like for instance XMAP215. This enzyme, in the absence of kinesin-8, attaches to and detaches from the MT tip with rates $k_\tn{on}$ and $k_\tn{off}$, respectively. Once bound, XMAP215 prevents kinesin-8 from reaching the tip, and processively polymerises the MT at rate $\eta_\tn{x}$, \emph{i.e.}\ the enzyme immediately binds to the newly formed tip site after polymerisation has occurred.
\end{enumerate}

We use the remainder of this chapter to give a concise summary of the results obtained recently for the \emph{neutral scenario}~\cite{Melbinger2012}: The combined effect of motor attachment in proximity of the minus-end (left boundary) and subsequent movement towards the plus-end (right boundary) leads to an accumulation of motors, which results in an antenna-like steady state profile~\cite{Varga2009,Reese2011}. At a certain distance from the minus-end the density profiles saturate to the equilibrium Langmuir density $\rho_\text{La}=K/(K+1)$~\cite{Leduc2012}. 
The resulting density profiles in vicinity of the minus-end is position-dependent, $\rho_-(x)$, and can be described by Lambert-$W$ functions~\cite{Parmeggiani2004}. Moving further towards the MT plus-end the density profile is determined by the interplay of motor current and the boundary conditions at the plus-end, which gives rise to a particular tip density $\rho_+(L)$. In a mean-field description~\cite{Chou2011}, this determines the length dynamics
\begin{equation}
\partial_t L (t)=-\delta\rho_+(L)+\gamma\,.
\label{eq:dtL}
\end{equation}
Steady state is reached at a critical density $\rho_+^c={\gamma}/{\delta}$, where $\partial_t L(t)=0$. 
Depending on whether the tip density $\rho_+(L)$ is smaller or larger than $\rho_+^c$ the MT grows or shrinks.

Because the motor current to the tip depends on the accumulation of motors along the MT, $\rho_-(x)$, the tip density depends on the actual length $L(t)$ of the MT. As a consequence a mechanism for MT regulation emerges: On a short MT, when the accumulation of motor density is low, also the tip density is low and the MT grows because the tip density lies below the critical threshold density, $\rho_+(L)<\rho_+^c$. 
This is in contrast to the case of a long MT where a higher density of motors accumulates along the MT and also the tip density is higher. Once the tip density exceeds the critical threshold value $\rho_+(L)>\rho_+^c$ the MT depolymerises.

\begin{figure*}[th]
\centering
\includegraphics[scale=1.0]{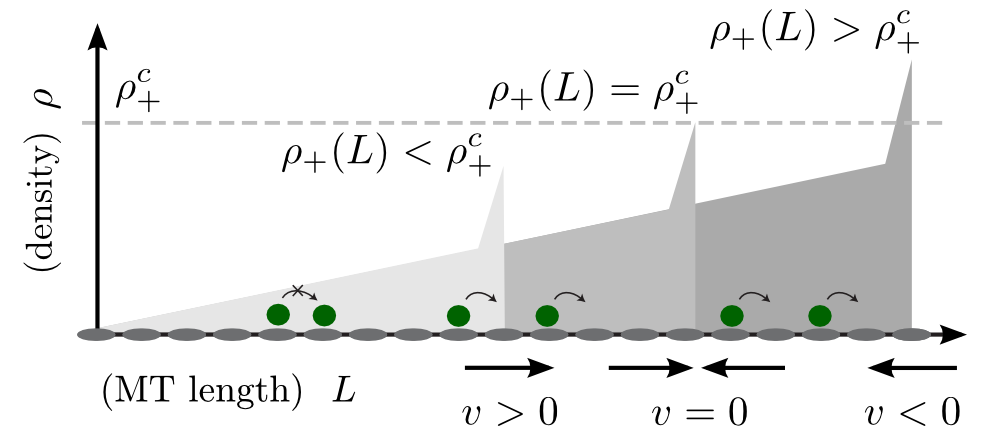}
\caption{Illustration of a linear motor density profile (shaded areas) and the threshold density $\rho_+^c$ (dashed line) for MT regulation.
Low tip density $\rho_+(L)$ ensues a growing MT, and a high tip density results in a shrinking MT. Note that the density at the tip generally has a spike-like shape~\cite{Pierobon2006}.\label{fig:cartoon_length}}
\end{figure*}

Figure~\ref{fig:cartoon_length} illustrates this mechanism. Shaded areas indicate density profiles for MTs of different length and also schematically account for the fact that the tip density is length-dependent and has a spike like shape. The dashed line shows a threshold value for the tip density, above and below which the MT shrinks and grows respectively. 

\begin{figure*}[t]
\centering
\includegraphics[scale=1.0]{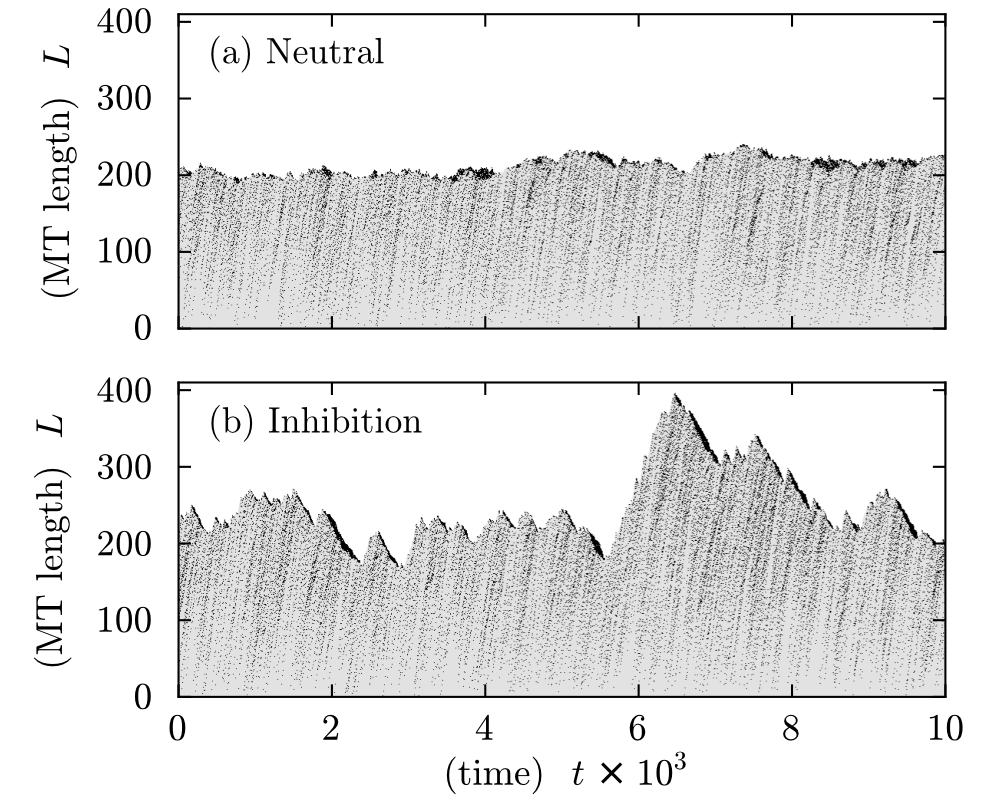}
\caption{
Kymographs of how molecular motors regulate a MT. In the neutral case (a) the system displays a higher accuracy in length regulation ($\delta=0.2$, $\gamma=0.1316$). This is in contrast to the case of growth inhibition (b) where the system displays intermittent dynamics ($\delta=0.2$, $\eta=0.385$).
A close look reveals different patterns of motor accumulation at the MT tips in the two scenarios. 
Attachment and detachment rates are $c \, \omega_\tn{on}=0.001$ and $\omega_\tn{off}=0.003$.\label{fig:trajectories}\label{fig:phen}}
\end{figure*}
\section{Motor and Microtubule Dynamics}\label{sec:dynamics}\vspace{-.3cm}
Though at first sight the \emph{neutral} and the \emph{inhibition} scenario as introduced above appear very similar, there are actually strong qualitative differences in the ensuing length dynamics. Fig.~\ref{fig:phen}(a) and (b) show kymographs for the neutral and the inhibition scenario, respectively, as obtained from stochastic simulations employing Gillespie's algorithm~\cite{Gillespie2007}. While in the neutral scenario the overall length of the MT stays approximately constant with only small fluctuations, the length dynamics for the inhibition scenario is intermittent with extended episodes of filament growth and shrinkage, reminiscent of the dynamic instability~\cite{Mitchison1984}. Note that for the inhibition scenario there is significant accumulation of motors at the MT tip during periods of depolymerisation. 

To understand how the system alternates between periods of growth and shrinkage, let us turn to a mathematical description of the dynamics. As already noted in the previous section, the length change of the MT is determined by the tip density $\rho_+$, \emph{i.e.} the probability that the MT tip is occupied by a molecular motor, 
\bequ
\partial_t L =
\begin{cases}
 -\delta \rho_+ + \gamma & \text{(neutral scenario)}\, , \\
 -\delta \rho_+ + \eta(1-\rho_+) & \text{(inhibition scenario)}\,.
\end{cases}
\label{eqn:drift}
\eequ
Here the first term on the right stands for depolymerisation, and the second term describes polymerisation dynamics of the neutral and the inhibition scenario, respectively. Equation~\eqref{eqn:drift} shows that depending on the magnitude of the tip density, $\rho_+$, the MT either grows or shrinks: For large tip densities, depolymerisation is strong and the MT shrinks, while the MT grows for small tip densities; see Fig.~\ref{fig:cartoon_length}. The critical tip densities, $\rho_+^c$, where the filament length becomes stationary read:
\bequ
\rho_{+}^c= \begin{cases}
\frac{\gamma}{\delta} & \text{(neutral scenario)}\, , \\
\frac{\eta}{\delta+\eta} & \text{(inhibition scenario)}\,.
\end{cases}
\label{eqn:rhoc_ex}
\eequ
To make further progress one needs to determine the actual tip densities employing a mean-field approach for the motor dynamics along the MT~\cite{Melbinger2012}.

\begin{table}[t]
\centering
\setlength{\extrarowheight}{2.pt}
\begin{tabular*}{8.4cm}{@{\extracolsep{\fill}}lll}
\hline \hline
{\bf Kinesin-8}  & model & experiment~\cite{Varga2009}\\
\hline 
speed &$\nu=1$& $6.35$ steps/sec\\
attachment &$\omega_\tn{on}$& $ 24\, (\tn{nM}\, \tn{min}\, \mu\tn{m})^{-1}$\\
detachment & $\omega_\tn{off}$ & $4.8 \cdot 10^{-3}\,\tn{sec}^{-1}$\\
depolymerisation&$\delta$&n/k~ \cite{Reese2011}\\
tip-detachment$\dagger$&$\beta$& $0.1-0.01\,\tn{sec}^{-1}$\\
\hline \hline
{\bf MT growth}& model& experiment$\ddagger$\\
\hline
neutral  &$\gamma$ & n/k\\
inhibition  & $\eta$ & n/k\\
\hline \hline
{\bf XMAP215} & model& experiment~\cite{Brouhard2008}\\
\hline 
attachment & $k_\tn{on}$& $ 0.1\, (\tn{nM}\, \tn{sec}\, \mu\tn{m})^{-1}$\\
detachment &$k_\tn{off}$& $ 3.8\,\tn{sec}^{-1}$\\
polymerisation &$\eta_\tn{x}$& $6.6\, \tn{dimers}/\tn{sec}$ \\
\hline 
\end{tabular*}
\caption{Quantification of model parameters for kinesin-8 and XMAP215. 
$\dagger$~Tip-detachment rates of different kinesin-8 constructs are: $10-55\,\tn{sec}$~\cite{Stumpff2011}; $20-40\,\tn{sec}$~\cite{Su2011}; $80\,\tn{sec}$~\cite{Varga2009}. In Ref.~\cite{Varga2009} it is shown that dwell times at the tip depend on motor concentration, suggesting cooperative effects of motors at the tip. A theoretical analysis is given in Ref.~\cite{Reese2011}.
$\ddagger$~{MT growth speeds in the presence of kinesin-8s in vivo are $1.3\, \mu\tn{m}/\tn{min}$~\cite{Gupta2006}; $2\, \mu\tn{m}/\tn{min}$~\cite{Stumpff2008,Tischer2009}. 
Rate constants of individual growth events, however, are not available to our knowledge and the complexity of the process~\cite{Gardner2011} renders it difficult to quantify the damping effects of kinesin-8~\cite{Du2010}.}\label{tab:rates}}
\end{table}

\subsection{Phase behaviour and tip densities}\label{sec:simplified}
\begin{figure*}[t]
\centering
\includegraphics[scale=1.0]{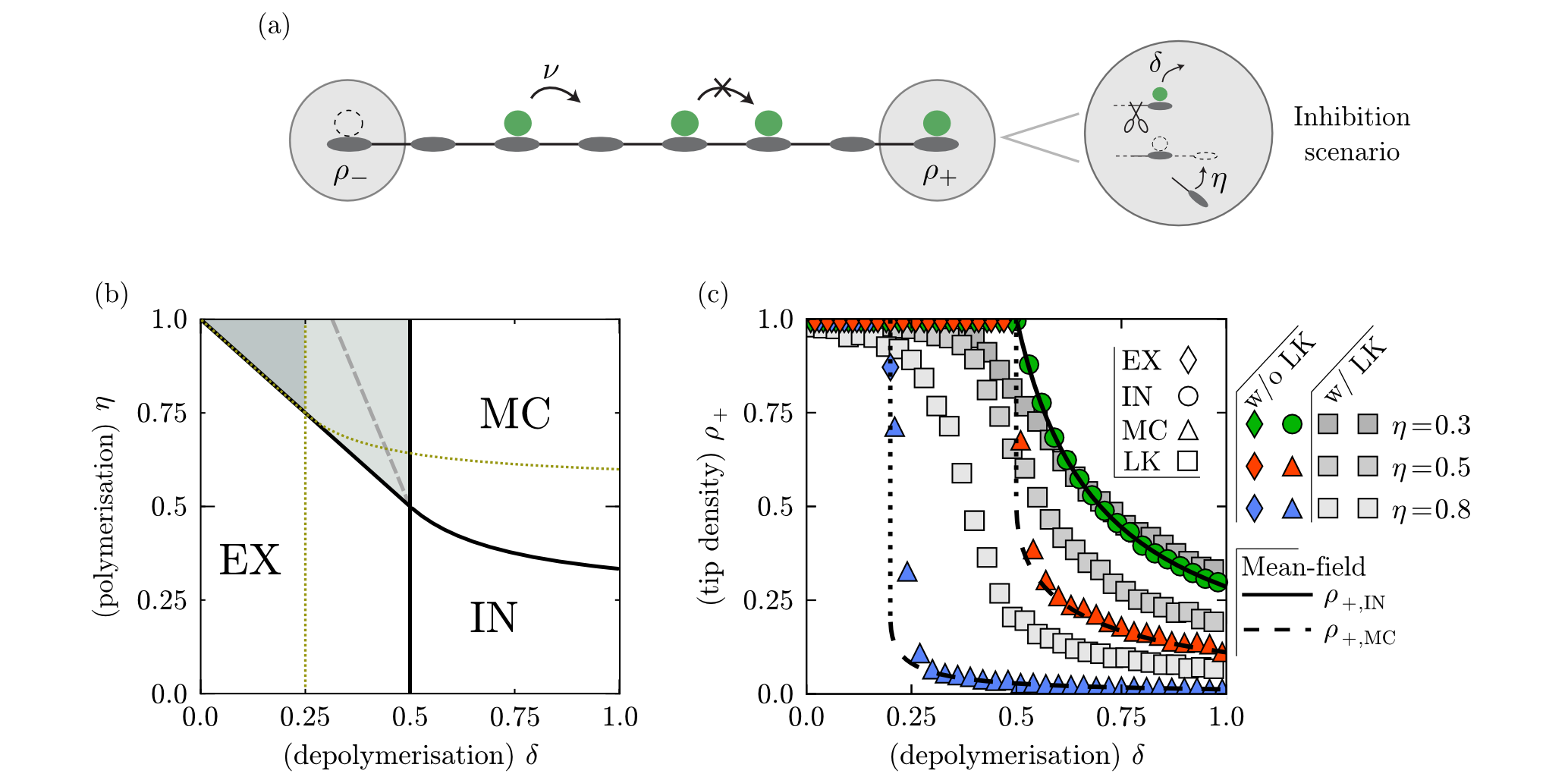}
\caption{
(a) Illustration of the \emph{simplified model} with a constant particle reservoir $\rho_-$ at the minus-end, and where Langmuir kinetics (LK) is not accounted for.
(b) Mean-field phase diagram for the simplified model (inhibition scenario) for two different values of the reservoir density: $\rho_-=0.5$ (solid black), and $\rho_-=0.25$ (dotted). Dashed line indicates the phase boundary obtained from stochastic simulations of the simplified model including LK with on- and off-rates $c\omega_\text{on}=\omega_\text{off}=0.005$. 
(c) Mean-field solutions for tip densities at various growth rates $\eta$ indicated in the graph compared to simulation data with and without LK. 
Different phases (IN/EX/MC) are indicated by symbols and lines and refer to analytic results; cf. Eqs.~\eqref{eqn:exclusive_IN}, and \eqref{eqn:exclusive_MC}. The dotted line indicates a discontinuous transition between the EX and MC phase. The lattice was initiated with random configurations of motors with bulk density $\rho_\tn{b}=0.5$.\label{fig:simplified}\label{fig:phasediagram}\label{fig:tip_density}}
\end{figure*}

\vspace{-.3cm}
For biologically relevant parameter ranges, the time scales of the tip dynamics and the motor dynamics are comparable, cf. Table~\ref{tab:rates}. Therefore, the motor density profile quickly adapts to changes in the tip density and one can readily assume that the tip density and the bulk density are adiabatically coupled~\cite{Melbinger2012}. Moreover, experimental data also show that both the attachment and the detachment rates, $\omega_\tn{on}$ and $\omega_\tn{off}$, are very small \cite{Varga2009}. This suggest to consider a \emph{simplified model} where one neglects the attachment and detachment kinetics, and assumes that a constant density $\rho_-$ serves as a particle reservoir at the left end of a lattice with fixed size $L$; see Fig.~\ref{fig:simplified}. This allows us to focus on the dynamics at the plus-end and unravel how it depends on the reservoir density $\rho_-$. Due to the adiabatic coupling between boundary and bulk, the results for the full model can be inferred from the simplified model upon replacing $\rho_-$ by the actual spatially varying profile $\rho_-(x)$. 

Since there is particle conservation, the dynamics of the tip density is given by 
\bequ
\partial_t \rho_+ =J_\text{b} (\rho_b, \rho_+) -J_\text{exit} (\rho_+)\, ,
\label{eqn:tip_dynamics}
\eequ
where $J_\text{b}$ and $J_\text{exit} = \delta \rho_+$ denote the bulk current and the loss rate of motors due to depolymerisation, respectively. Calculations are conveniently performed in a frame comoving with the MT tip. Then the bulk currents for the neutral (N) and the inhibition (I) scenario read in a mean-field approximation~\cite{Melbinger2012}:
\bse\bequa
J_\text{b}^\text{N} 
&=& \rho_\text{b}(1-\rho_\text{b})-\gamma\rho_\tn{b}+\delta \rho_+ \rho_\text{b}
\label{current_nonex}    
\, ,\\
J_\text{b}^\text{I} 
&=& \rho_\text{b}(1-\rho_\text{b})-\eta  \rho_\text{b} (1-\rho_+)+\delta \rho_{+} \rho_\text{b}
\, .
\label{current_ex}
\eequa\ese
Here $\rho_b$ denotes the motor density in the bulk of the MT, and the first term describes the current due to hopping processes accounting for particle exclusion on neighbouring sites. The remainder of the terms indicate polymerisation and depolymerisation currents, which in a comoving frame simply correspond to simultaneous movement of all particles on the MT lattice to the left and right end, respectively. 

The stationary state of the model is determined by a balance of currents, or, in other words, the fixed point of Eq.~\ref{eqn:tip_dynamics}:~$\rho_+ \delta = J_\text{b} (\rho_b, \rho_+)$. Solving for the tip density one finds
\bequ
\rho_{+} = \rho_{+}(\rho_\tn{b},\delta,\eta)=\frac{\rho_\tn{b} (1-\eta -\rho_\tn{b})}{\delta (1-\rho_\tn{b})-\eta \rho_\tn{b}}\, ,\label{eqn:tip_exclusive}
\eequ
for the inhibition scenario. 
The tip density is determined by the bulk particle flux towards the tip, and, at the same time, the bulk density depends on the molecular processes at the MT tip. To make progress with the analytical calculations, it is necessary to have some phenomenological knowledge about the nature of the density profiles and their stability with respect to fluctuations. For exclusion processes, there are in general three distinct phases, each of which corresponds to different bulk densities $\rho_\tn{b}$ and ensuing bulk currents~\cite{Derrida1992,Schuetz1993}:
\begin{itemize}
\item {\it IN phase}: In this phase the particle current that enters the system at the minus-end determines the bulk density. For TASEP this phase is also called \emph{low density phase}.
\item {\it EX phase}: The bulk density is determined by the current of particles that leave the system at the right boundary (TASEP: \emph{high density phase}). 
\item {\it MC phase}: In this phase the \emph{maximal current} (MC) through the system determines the bulk density. It corresponds to a local maximum in the current density relation $J_\text{b} (\rho_\text{b})$. In contrast to the two other phases the bulk density in the MC phase is independent of the boundary conditions.
\end{itemize}
Moreover, for exclusion processes, there are two possibilities to account for the boundary conditions at the left and right end. Either there is a domain wall (DW) delineating a low density region, $\rho_-$, from a high density region, $\rho_+$, or there are boundary layers~\cite{Hager2001} at one of the MT ends.

\subsubsection{Density perturbations and domain wall theory}
\label{sec:dwtheory}
\vspace{-.3cm}
To make progress on the phase diagram we need to investigate the stabilities of the aforementioned DW and bulk density. To this end we introduce two important criteria that allow to analyse the stability of perturbations in exclusion processes known as DW-theory and the extremal current principle (ECP)~\cite{Krug1991,Kolomeisky1998,Popkov1999}.
First we consider the stability of a bulk density $\rho_\tn{b}$ against a density perturbation. Such a perturbation travels at the collective velocity~\cite{Krug1991,Kolomeisky1998}
\begin{equation}
u_\text{coll}=\partial_{\rho} J_\tn{b} (\rho, \rho_+) \mid_{\rho= \rho_\tn{b}}\, .
\end{equation}
Since for $u_\tn{coll} < 0$ density perturbations move towards the minus-end, they do not affect the tip density and thereby the EX phase remains stable. In contrast, for $u_\tn{coll} > 0$, perturbations move towards the plus-end which renders the IN phase stable against density fluctuations. Note that the collective velocity $u_\tn{coll} = 0$ in the MC phase (by definition). Second, we consider the stability of DWs. A DW between a left $\rho^\text{left}$ and right density $\rho^\text{right}$ travels at a velocity,
\begin{align}
u_{\tn{DW}}=\frac{J_\tn{b}(\rho^\text{left}, \rho_+)-J_\tn{b}(\rho^\text{right}, \rho_+)}{\rho^\text{left}-\rho^\text{right}}\, .
\label{eq:vshock}
\end{align} 
Depending on the sign of this velocity the phase corresponding to $\rho^\text{left}$ or $\rho^\text{right}$ is stable~\cite{Kolomeisky1998}. Taken together $u_\tn{coll}$ and $u_\tn{DW}$ lead to analytic results for bulk and tip densities in the various phases; see Table~\ref{tab:exact}.

\subsubsection{Phase diagram for the inhibition scenario}\label{sec:phasediagram}
\vspace{-.3cm}
With the methods introduced in the previous section it is a straightforward task to derive the densities and the ensuing phase behaviour of the simplified model. Since the neutral scenario has already been discussed previously~\cite{Melbinger2012}, we here focus on the inhibition scenario. In the IN phase the bulk density is (by definition) given by the reservoir density at the left boundary: $\rho_\text{b}^\text{IN}=\rho_-$. With the stationarity condition, Eq.~\eqref{eqn:tip_exclusive}, one finds that the tip density is a function of the reservoir density
\bequ
\rho_{+}^{\tn{IN}}(\rho_-,\delta,\eta)=\frac{\rho_- (1-\eta -\rho_-)}{\delta  (1-\rho_-)-\eta \rho_-}\, .\label{eqn:exclusive_IN}
\eequ
Note, however, that this is a stable solution of Eq.~\ref{eqn:tip_dynamics} only outside of the shaded area indicated in the phase diagram shown in Fig.~\ref{fig:phasediagram}(b). In the EX phase, the bulk density is given by the right boundary, $\rho_\text{b}^\text{EX}=\rho_+$, and Eq.~\eqref{eqn:tip_exclusive} leads to the striking result that the MT tip is \emph{always} occupied by a molecular motor,
\bequ
\rho_{+}^{\tn{EX}}=1 \,, \label{eqn:exclusive_EX}
\eequ
in stark contrast to the corresponding result in the neutral scenario; see Table~\ref{tab:exact}. It implies that a MT always depolymerises for those parameter regimes where the system is in the EX phase. Similar as in Ref.~\cite{Reese2011} we attribute this behaviour to the slow depolymerisation rate in the EX phase, $\delta < \rho_-$. It implies that motors leave the tip more slowly than they arrive. Then the MT tip acts as a bottleneck for molecular transport and induces a traffic jam with $\rho_{+}^{\tn{EX}}=1$ at the plus-end. For the MC phase, the bulk density is given by the maximum of the bulk current $J_\tn{b}^\tn{I}$: 
\bequ
\rho_{\text{b}}^\text{{MC}}=\frac{\delta -\sqrt{\delta  \eta  (\delta +\eta -1)}}{\delta +\eta }\, .
\eequ
Upon using this bulk density in Eq.~\eqref{eqn:tip_exclusive} gives a constant value for the tip density in the MC phase which is independent of the reservoir density
\bequ
\rho_{+}^{\tn{MC}}=\frac{\delta+\eta  (\eta +\delta -1)-2 \sqrt{\eta  \delta  (\eta +\delta -1)} }{(\delta+\eta)^2}\, .\label{eqn:exclusive_MC}
\eequ

Knowing the tip densities, we can now use the domain wall theory explained above (see section \ref{sec:dwtheory}) to determine the transition lines between the various phases. The DW velocity gives the direction in which a DW between two densities, one from the left and one from the right, travels. To employ this criteria, we first have to identify the respective densities. Let us start with $\rho_\text{left}$: The density at the minus-end is in general determined by the entering current, corresponding to a tip density $\rho^+_{\text{IN}}$, Eq.~\ref{eqn:exclusive_IN}. This tip density, however, is only stable against small perturbations if $u_\text{coll}\geq0$. For parameters where $u_\text{coll}<0$ the density from the left is decreased to $\rho_\text{left}=\rho^+_{\text{MC}}$. This sign-change of the collective velocity defines the phase boundary between the IN and MC phase: $\eta = \delta(\rho_- -1)^2/(\delta -\rho_-^2)$. Taken together, the density on the left of the DW is given by $\rho_\text{left}=\text{Min}[\rho^+_{\text{IN}},\rho^+_{\text{MC}}]$. The density at the right of the DW, $\rho_\text{right}$, is determined analogously. Since in that regime the collective velocity is strictly negative we simply have $\rho_\text{right}=\rho^+_{\text{EX}}=1$. 
Upon using the above expressions for $\rho_\text{left}$ and $\rho_\text{right}$ in Eq.~\eqref{eq:vshock} gives the remaining phase boundaries: With $\rho^\text{left}=\rho_-$, $\rho^\text{right}=1$, and $\rho_+=1$ one obtains $u_{\tn{DW}}= \delta - \rho_-$, implying that the phase boundary between the IN and EX phase is given by $\delta = \rho_-$. The boundary line $\delta + \eta = 1$ signifies that above this line the stationary solution given by Eq.~\ref{eqn:exclusive_IN} becomes unstable. This instability gives rise to interesting motor dynamics, in particular, a subtle dependence of the ensuing stationary profile on the initial condition. While these effects are certainly worthwhile studying they are irrelevant for our main focus, namely MT regulation, and, hence, we refrain from further analysing this regime here.

Taken together, the above analysis gives the phase diagram shown in Fig.~\ref{fig:phasediagram}(b) for two different values of the reservoir density $\rho_-$.
The general trend is that with decreasing reservoir density the parameter domain where the IN phase is stable expands. 

\begin{table*}[t]
\setlength{\extrarowheight}{2.pt}
\begin{tabular*}{172.5mm}{@{\extracolsep{\fill}}lllll}
\hline \hline
\bf Tip density\\
\hline 
Inhibition & 
$\rho_{+}^{\tn{IN}}=\frac{{\rho_-  (1-\eta -\rho_- )}}{{\delta-\rho_-  (\eta +\delta ) }}$&
$\rho_{+}^{\tn{EX}}=1$&
$\rho_{+}^{\tn{MC}}=\frac{\eta  (\eta +\delta -1)+\delta -2 \sqrt{\delta\eta  (\eta +\delta -1)} }{{(\eta +\delta )^2}}$\\
\hline
Neutral~\cite{Melbinger2012} &
$\rho_{+}^{\tn{IN}}=\frac{\rho_-  (1-\gamma -\rho_- )}{\delta (1- \rho_-)}$&
$\rho_{+}^{\tn{EX}}=1-\frac{\gamma }{1-\delta }$&
$\rho_{+}^{\tn{MC}}=\frac{1-\sqrt{\gamma} }{\delta}$\\
\\
\hline \hline
\bf Critical growth\\
\hline 
Inhibition& 
$\eta_c^\tn{IN}=\frac{\delta   \rho_-(1-\rho_- ) }{\delta - \rho_-(1-\rho_- ) }$ &
$\eta_c^\tn{EX}=1-\delta$ [see Eq.~\eqref{rho_c_EX}]&
$\eta_c^\tn{MC}=\frac{\delta }{4 \delta -1}$\\
\hline
Neutral~\cite{Melbinger2012}& 
$\gamma_c^\tn{IN}=\rho_-(1 -  \rho_-)$ &
$\gamma_c^\tn{EX}=\delta  (1 -\delta )$ &
$\gamma_c^\tn{MC}=\frac{1}{4}$\\
\hline
\end{tabular*}
\caption{Analytic results for the tip densities $\rho_+$ in the different phases IN/EX/MC and the critical growth rates $\eta_c$ and $\gamma_c$ for the inhibition and neutral scenario, respectively. Note that $\eta_c^\tn{EX}$ is obtained from the phase boundary of the EX phase as derived in the main text.\label{tab:exact}}
\end{table*}

The analytical results obtained from mean-field theory agree nicely with the stochastic simulations, see Fig.~\ref{fig:phasediagram}(c), in case LK is neglected. For a depolymerisation rate $\eta = 0.3$, concomitant with the phase transition from the IN to the EX phase, the tip density increases upon lowering the depolymerisation rate $\delta$ and then continuously saturates at $\rho_+=1$ as the EX phase is reached. In contrast, for $\eta \geq 0.5$, there is a \emph{discontinuous jump} in the tip density as one passes from the MC into the EX phase; see discussion above. 

The stochastic simulations with LK show a quite significant increase in the magnitude of the tip density in the MC phase, in particular in the shaded area of the phase diagram, Fig.~\ref{fig:phasediagram}(b). 
We attribute this to the fact that the Langmuir density in bulk, $\rho_\text{La}$, acts as a source for kinesin-8 motors which tends to increase the motor density on the MT and at the tip.
Though these effects are interesting and worthwhile studying they are not important for our main concern here, namely regulation of MT length. 
As discussed previously~\cite{Melbinger2012} and elaborated on later in section \ref{sec:length}, MT regulation is possible only if the density profile is determined by the particle current at the minus-end, \emph{i.e.}\ if the system in its stationary state is in the IN phase. 
In that case, even adding LK in the simulations has only a minor effect on the magnitude of the tip density, and we can savely use the analytical mean-field results to further analyse the stationary MT length.
\begin{figure*}[t]
\centering
\includegraphics[scale=1.0]{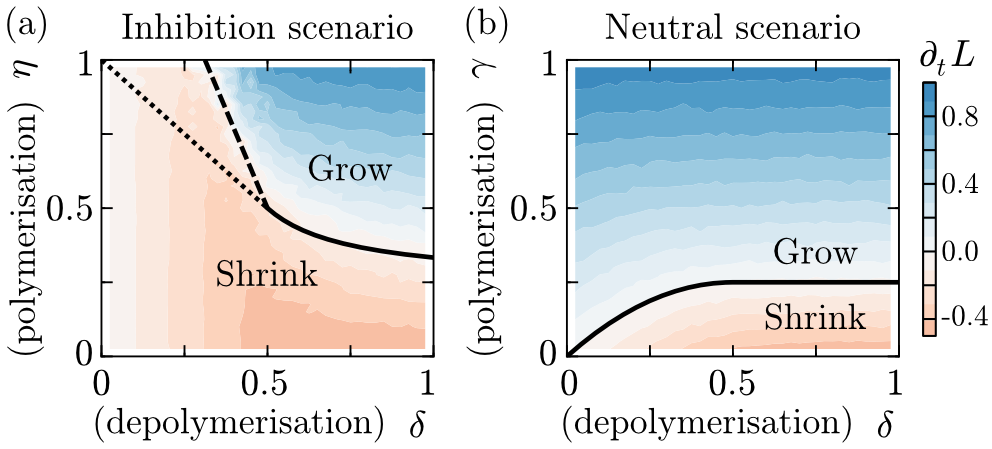}
\caption{Drift velocity of the MT tip, $v=\partial_t L$, as a function of the polymerisation and depolymerisation rates for the \emph{inhibition} (a) and the \emph{neutral} (b) scenario obtained from stochastic simulations for the simplified model with LK; colour code indicates the magnitude of the drift. Solid lines indicate where the MT velocity is zero, $\eta_c (\delta)$ and $\gamma_c (\delta)$, as obtained from the analytical calculations; see Table~\ref{tab:exact}. The dotted line is obtained from the analytical theory; it coincides with the boundary line $v=0$ and agrees well with numerical simulations where LK has been turned off. The dashed line is the numerically determined boundary $\eta_c$ when LK is turned on. Stochastic simulations including LK were performed with motor attachment and detachment rates $c\omega_\text{on}=\omega_\text{off}=0.005$, respectively, $\rho_-=0.5$ and system size $200$.
\label{fig:drift}\label{fig:comparison}}
\end{figure*}
\subsection{Dynamics of the microtubule length}\label{sec:speed}\vspace{-.3cm}
Figure \ref{fig:drift} shows the results of our stochastic simulations with LK for the MT drift velocity, $v=\partial_t L$ as a function of the depolymerisation and the polymerisation rates for both the inhibition and the neutral scenario. 
There are well defined boundaries, $\eta_c (\delta, \rho_-)$  and $\gamma_c (\delta, \rho_-)$, separating regimes in which MTs grow and shrink, respectively.
Since the tip density, $\rho_+$, dictates MT dynamics, see Eq.~\eqref{eqn:drift}, those boundaries can be readily calculated upon comparing the tip densities listed in Table~\ref{tab:exact} with the critical tip density, Eq.~\eqref{eqn:rhoc_ex}. 
For the inhibition scenario we find that for $\delta < \rho_-$ the critical tip density coincides with the phase boundary of the EX phase
\bequ
\eta_c = 1 -\delta \, ,
\label{rho_c_EX}    
\eequ
while for $\delta > \rho_-$ it lies either within the MC or the IN phase:
\bequ
\eta_c = 
\begin{cases}
\frac{\delta \rho_-(1-\rho_- )}{\delta-\rho_-(1-\rho_-)}
& \text{for} \quad \rho_- < 1/2 \, ,\\
\frac{\delta/4 }{\delta -1/4} 
& \text{for} \quad \rho_- > 1/2 \, ;\\
\end{cases}
\label{rho_c_MC}
\eequ
see Table~\ref{tab:exact} for a summary together with the results for the neutral scenario. These analytical results are in perfect accordance with our stochastic simulations [Fig.~\ref{fig:drift}] with one interesting exception for the inhibition scenario, namely the boundary line of the EX phase for $\delta < \rho_-$. Since we recover agreement between stochastic simulations and analytical calculations by switching off LK in our stochastic simulations, we can fully attribute this difference to the effect of attachment and detachment of motors in bulk, as discussed in the previous section, \ref{sec:phasediagram}; cf. dotted and dashed lines in Fig.~\ref{fig:drift}(a). Further, the differences between both scenarios are significant, cf. Figs.~\ref{fig:drift}(a) and (b), respectively. In the inhibition scenario, the regime where MTs shrink -- and hence regulation becomes possible -- is much broader since kinesin-8 inhibits MT growth when bound to the tip: For small depolymerisation rate $\delta$, motors reside at the MT end for a relatively long time, which dramatically broadens the regime of MT shrinkage.

\begin{figure}[t]
\centering
\includegraphics[scale=1.0]{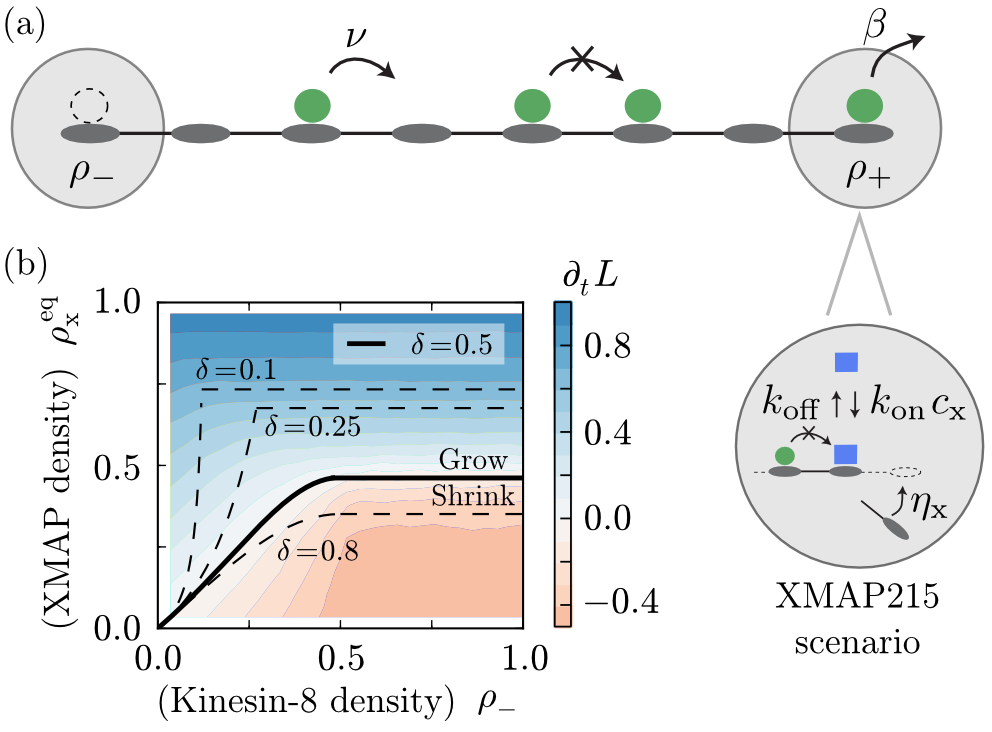}
\caption{Microtubule dynamics with kinesin-8 and XMAP215: (a) Simplified model with motor detachment from the tip (rate $\beta$) and tip-binding of XMAP215. (b) MT growth velocity as a function of kinesin-8 and XMAP215 density. 
A XMAP215 density of $\rho_\tn{x}^\tn{eq}=0.5$ corresponds to  $c_\tn{x}\approx 10\,\mu$M, and the concentration of kinesin-8 is approximately $c\approx 1.5\,\tn{nM}$ for a half-filled lattice $\rho_\tn{b}=0.5$, see Table~\ref{tab:rates}.
\label{fig:xmap_dynamics}}
\end{figure}

\subsection{Interplay between kinesin-8 and polymerase XMAP215}\label{sec:xmap}
\vspace{-.3cm}
In this section we compare the dynamics of the inhibition scenario with a model which explicitly accounts for a second protein, XMAP215, that enzymatically facilitates MT growth; see Fig.~\ref{fig:xmap_dynamics}(a) and section \ref{sec:model}. 
Since XMAP215 and kinesin-8 mutually exclude each other at the MT tip, one expects strong similarities between those scenarios. In order to compare with an analytically tractable lattice gas model we performed the stochastic simulations for the simplified model without LK~\footnote{Note that in order to achieve a more realistic description of the tip related processes, we also include tip-detachment of kinesin-8 at a rate of $\beta=0.02$ as suggested by experiments cf. Table~\ref{tab:rates}.}. Figure~\ref{fig:xmap_dynamics} shows the regimes of MT growth and shrinkage as a function of kinesin-8 and XMAP215 densities for a set of depolymerisation rates $\delta$. The general trend is that the regime where MTs shrink is enlarged with smaller depolymerisation rates.

At the mean-field level, the equilibrium density of XMAP215 at the MT tip is given by the product $\rho_\tn{x}=\rho_\tn{x}^{\tn{eq}}(1-\rho_+)$, where $1-\rho_+$ is the probability that kinesin-8 is not bound and $\rho_\tn{x}^{\tn{eq}}$ denotes the Langmuir isotherm for XMAP215 binding:
\bequ
\rho_\tn{x}^{\tn{eq}}=\frac{c_\tn{x}k_\tn{on}}{c_\tn{x}k_\tn{on}+k_\tn{off}}\,.\label{eqn:xmap_density}  
\eequ
Here $c_\tn{x}$ is the XMAP215 concentration in solution, and $k_\tn{on}$ and $k_\tn{off}$ are the attachment and detachment rates of the enzyme to and from the MT tip, respectively. This mean-field approximation neglects that the presence of XMAP215 at the MT tip influences the current of kinesin-8 to the MT end because it could block the motor particles~\cite{Wood2009}. 
Fortunately, since the polymerisation rate of XMAP215, $\eta_\tn{x}$, and the walking speed, $\nu$, of kinesin-8 are almost the same~\cite{Brouhard2008,Varga2009} the two molecules rarely interact. This implies that a model explicitly accounting for XMAP215 can be reduced to the inhibition scenario with an effective polymerisation rate given by
\begin{equation}
\eta= \eta_\tn{x} \, \rho_\tn{x}^{\tn{eq}}\, .
\label{eqn:xmap_poly} 
\end{equation}
Indeed, as can be inferred from Fig.~\ref{fig:xmap_dynamics}(b), the predictions of the effective inhibition scenario agree nicely with the numerical simulations. Taken together this implies that the inhibition scenario may serve as a minimal model to include other MT associated proteins that antagonise the depolymerisation activity of kinesin-8. 
\subsection{Microtubule regulation}\label{sec:length}
\begin{figure*}[t]
\centering
\includegraphics[scale=1.0]{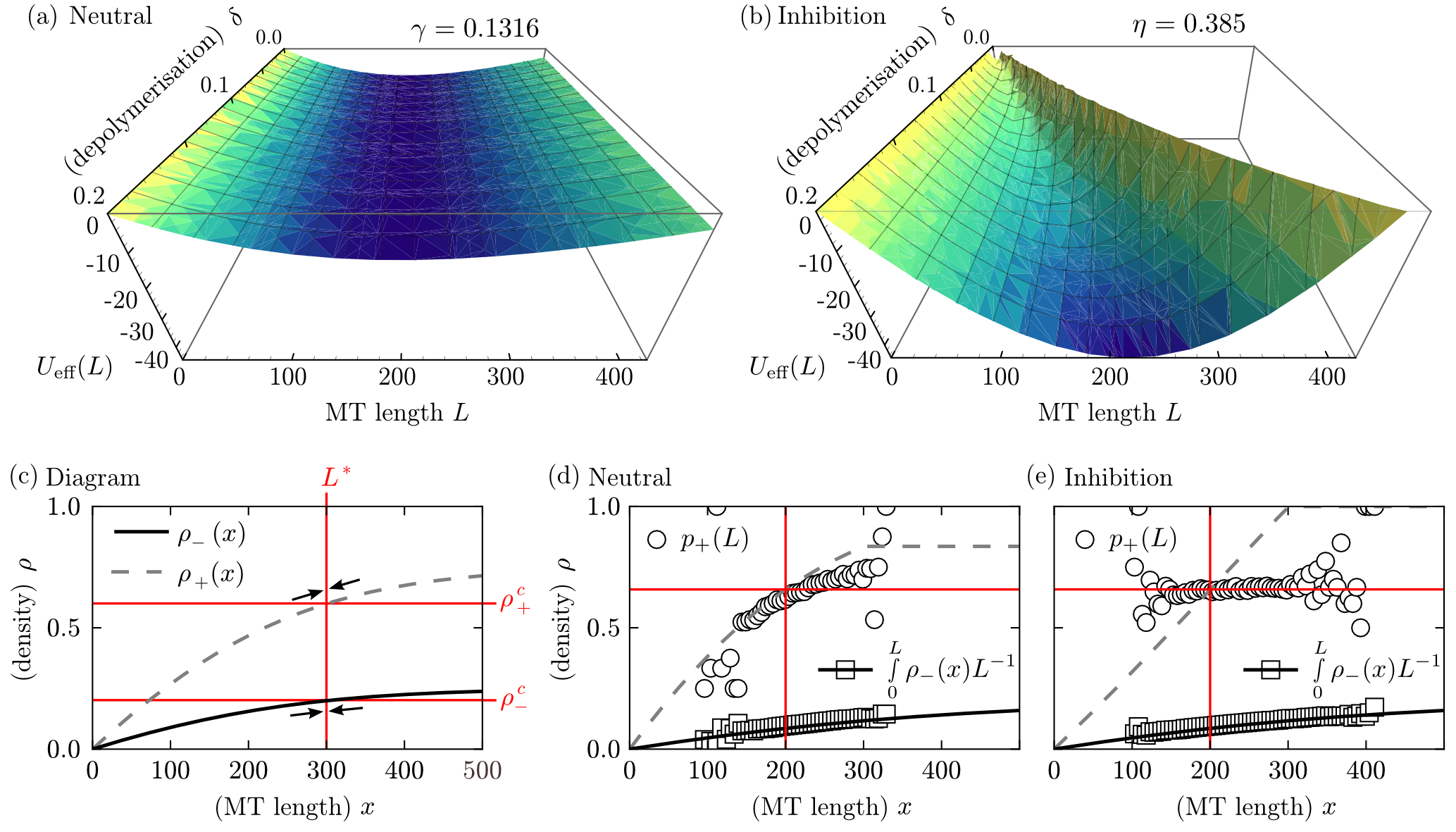}
\caption{Effective potentials for the inhibition and the neutral scenario, (a) and (b), for the trajectories shown in Fig.~\ref{fig:trajectories}. The diagram in (c) illustrates how threshold densities for the tip density $\rho_+(x)$ and minus-end density $\rho_-(x)$ are defined, and how both quantities together set MT length $L^*$. Panels (d) and (e) show data for the accumulated density, and the ensuing probability of tip occupation $p_+ (L)$. For the inhibition scenario $p_+ (L)$ is constant, while in the neutral scenario it is length-dependent and thus samples the effective potential, \emph{i.e.} $p_+ (L) = \rho_+ (L)$. 
\label{fig:potential}}
\end{figure*}

\vspace{-.3cm}
We now consider the full model for a MT of finite length $L$, where LK leads to an accumulation of kinesin-8 motors along the MT. As discussed in section~\ref{sec:model}, the ensuing antenna-like profile $\rho_-(x)$ can be calculated within the framework of the TASEP/LK model~\cite{Parmeggiani2003,Parmeggiani2004}; these theoretically predicted profiles have recently been confirmed by in vitro experiments~\cite{Leduc2012}. Now length regulation becomes possible if this spatially varying profile translates into a length-dependent velocity $v(L)$ of the MT tip~\cite{Howard2007,Melbinger2012}. This requires that the tip density $\rho_+ (L)$ depends on $\rho_-(L)$ which is the case only for the IN phase; see Eq.~\ref{eqn:exclusive_IN}. Then the tip density reads
\bequ 
\rho_{+} (L) 
= \rho_{+}^{\tn{IN}} (\rho_-(L),\delta,\eta) \, .
\label{eqn:rhoplusL}
\eequ
Upon inserting the ensuing length-dependent tip density into Eq.~\ref{eqn:drift} one obtains a length-dependent velocity $v(L)$. It is instructive to define an \emph{effective potential}
\bequ 
U_\text{eff} (L) = - \int_0^L dx \; v(x) \, ,
\label{eqn:rhoplusL}
\eequ
whose minimum defines the stationary MT length $L^*$
\bequ 
\rho_{+} (L^*) = \rho_{+}^{\tn{IN}} (\rho_-(L^*),\delta,\eta) =\rho_+^c\, ,\label{eqn:Lstar}
\eequ
as illustrated in Figs.~\ref{fig:potential}(a-c). Tight length regulation is restricted to the regime where the critical density $\rho_-^c:= \rho_-(L^*)$ falls well into the linearly increasing antenna profile. The closer $\rho_-^c$ is to the Langmuir plateau $\rho_\tn{La}$ the less well defined is the stationary length; note that the effective spring coefficient 
\bequ
k(L):=U_\text{eff}^{''} (L) =
\begin{cases}
 \delta \, \rho'_+ (L)  & \text{(neutral scenario)} \\
 (\delta + \eta) \, \rho'_+ (L) & \text{(inhibition scenario)}
\end{cases}
\, , \label{eqn:spring_constant}
\eequ
is proportional to the slope of the profile, where prime denotes derivative; see also Fig.~\ref{fig:potential}(c). 

\begin{figure*}[t]
\centering
\includegraphics[scale=1.0]{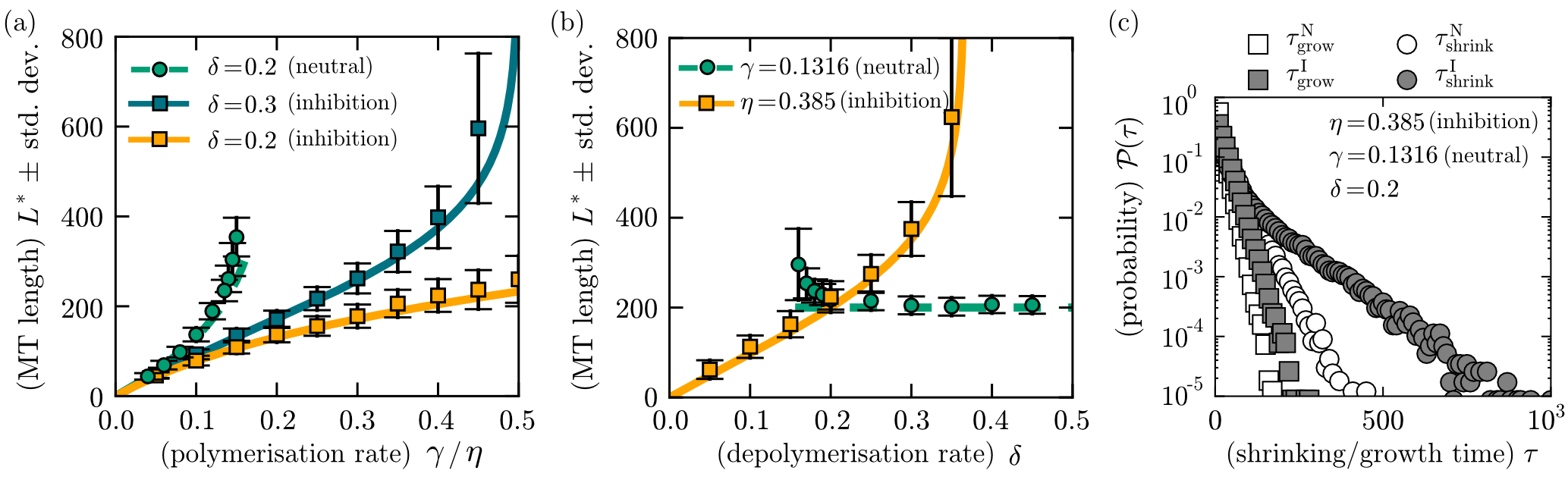}
\caption{Comparison of MT length $L^*$ for the inhibition (solid) and neutral (dashed) scenario with respect to polymerisation (a) and depolymerisation rates (b). 
The data was obtained single trajectories $L(t)$. Data points correspond to the most probable length of the process $L^*$; error bars denote the standard deviation of $L(t)$.
Motor attachment and detachment rates are $c \, \omega_\text{on}=0.001$ and $\omega_\text{off}=0.003$. In (c) probability distributions of the times, during which MTs shrink and grow, are shown for parameter values as in Fig.~\ref{fig:phen}. The exponential tails of the distributions support the view that the inhibition scenario follows dichotomous switching dynamics (see main text for details).  
\label{fig:length}}
\end{figure*}

As can be inferred from Figs.~\ref{fig:length}(a,b) the stochastic simulations agree nicely with the above analytical results for the stationary MT length $L^*$ in both scenarios, neutral and inhibition. Previous studies~\cite{Melbinger2012} have shown that the variance of the length can be obtained well upon using a van Kampen expansion for the stochastic dynamics of the MT length $L(t)$, which assumes that the tip density is adiabatically coupled to the motor density along the MT. 
This essentially amounts to saying that the MT length performs a random walk in the effective potential $U_\text{eff}(L)$. Such a picture is fully consistent with results obtained from our stochastic simulations: The observed stochastic trajectories resemble those of random walks in confinement; see Fig.~\ref{fig:phen}(a). 
More importantly, the numerically observed value for the probability that the MT tip is occupied, $p_+(L)$ agrees well with the mean-field tip density $\rho_+(x)$; see Fig.~\ref{fig:potential}(d). 
This implies that the stochastic trajectory samples the values of MT length $L(t)$ with a statistical weight determined by the effective potential $U_\text{eff} (L)$. Surprisingly, as can be inferred from Fig.~\ref{fig:potential}(e), this is not the case for the inhibition model which immediately invalidates a description of the stochastic dynamics in terms of a continuous random walk in the potential landscape shown in Fig.~\ref{fig:potential}(b). 
The latter would actually give rise to stochastic trajectories strongly confined to the stationary value $L^*$. In contrast, the actual stochastic trajectories for the inhibition scenario shown in Fig.~\ref{fig:phen}(b) rather resemble an intermittent dynamics similar to the behaviour of MT dynamic instability with abrupt transitions between growing and shrinking states~\cite{Mitchison1984}.

The key to understand this anomalous dynamics lies in realising that the stochastic length dynamics in the inhibition scenario is a \emph{dichotomous process} with only two states: While, if the MT tip is empty, the MT grows with a rate $\eta$, it shrinks with a rate $\delta$ if the MT tip is occupied by a kinesin-8 protein. 
In other words, depending on whether the MT tip is occupied or not, it is either in a shrinking or a growing state, respectively~\cite{Dogterom1993}. 
Consider a configuration where the tip is empty, and, hence, the MT is in a growing state (with average speed $\eta$). Then it will remain in this state for some time $\tau_\tn{grow}$ until the motor closest to the tip actually reaches the tip. Figure~\ref{fig:length}(c) shows the probability distribution of $\tau_\tn{grow}$ for the same parameters as in Figure~\ref{fig:length}(b). 
The distribution is clearly exponential with a typical time of the order $\sim 23 / \nu$. On the other hand, if the MT tip is occupied by a kinesin-8 protein, it will remain in this state and not depolymerise the tip for a time of the order of $\delta^{-1}$. During this time the filament does neither grow nor shrink, and the kinesin-8 protein at the MT tip acts as a strict bottleneck. 
As a consequence, an extended traffic jam may emerge at the MT tip by motors queuing up behind this bottleneck. These traffic jams can be clearly seen in Fig.~\ref{fig:phen}(b) as black clusters. 
The formation of such clusters is a nucleation process, and the duration of the shrinking state is determined by a subtle interplay between particles gained by stochastic arrival at the left end and depolymerisation dynamics. Interestingly, the probability distribution of $\tau_\tn{shrink}$ shows two typical time scales, and, in particular, a broad exponential tail with a typical time of the order $\sim 112 / \nu$. 
We leave a more detailed investigation of these interesting stochastic effects for future work. The main results we would like to emphasize here are that we have identified
two distinct time scales characteristic for prolonged growing and shrinking states. These time scales are macroscopic in the sense that they are much larger then the hopping time of individual motors (which we have set to $1$). This implies that also the typical lengths covered during the growing and the shrinking state are rather large; for the examples shown in Figure~\ref{fig:length}(c) they are on average approximately $8$ and $22$ lattice sites during polymerisation and depolymerisation, respectively. These large length scales explain why the probability to occupy the MT tip as obtained from the stochastic simulations is only weakly dependent on MT length. 

Taken together we find that in the neutral scenario MT length is tightly controlled. The variance in MT length is mainly determined by the width of the effective potential, or, equivalently by the effective spring coefficient $k(L) = \delta \, \rho'_+ (L)$. Hence, the slope of the antenna profile is the key determinant of length fluctuations. In contrast, for the inhibition scenario, extended periods of MT growth and shrinkage lead to large length fluctuations as can be seen from the kymographs in Fig.~\ref{fig:phen}. These large fluctuations ensue characteristic exponential tails in the filament's length distributions; the characteristic width of this distribution is shown as error bars in Fig.~\ref{fig:length}(b). Note also the different dependencies of the two models with respect to the depolymerisation rate. While, in the neutral scenario the length of the MT is independent of the depolymerisation rates $\delta$, it strongly affects MT length in the inhibition scenario.

\section{Discussion}
\label{sec:discussion}
\vspace{-.3cm}
We have analysed distinct molecular mechanisms of MT regulation by proteins which are able to catalyse growth and shrinkage of MTs. Specifically, our interest was in the interplay between kinesin-8 motors acting as depolymerases when bound to the MT tip and microtubule growth processes which are either spontaneous or also catalysed by proteins like XMAP215. We investigated two distinct scenarios: In a \emph{neutral scenario} MTs grow independently of whether a kinesin-8 motor is bound to the tip or not. In contrast, in an \emph{inhibition scenario}, the MT only grows if the MT tip is not occupied by a depolymerase. Experiments with a microtubule polymerising enzyme, XMAP215~\cite{Brouhard2008}, suggest a high binding rate for the microtubule tip through facilitated diffusion. Then, to a first approximation, one may model XMAP215 as a tip-binding protein which excludes binding of kinesin-8. As we have shown, this tip site exclusion leads to a dynamics which is equivalent to the inhibition scenario. 

The results obtained here show how interactions between individual proteins and the microtubule tip play an important role for microtubule regulation. There are three main findings: 
(i) Microtubule regulation is directly affected by motor traffic. It is influenced by the microtubule growth rate, and the attachment and detachment kinetics of motors to and from the microtubule.
Both parameters can be tuned in experiments through the tubulin concentration and the motor and salt concentrations~\cite{Leduc2012}, respectively.
(ii) Regimes of microtubule growth and shrinkage critically depend on the probability that a kinesin-8 motor is bound to the microtubule tip. 
(iii) Protein-microtubule interactions at the microtubule tip are key to distinguish different mechanism of microtubule regulation, like for example intermittent dynamics or tight length control. 

The parameter regimes, where motor traffic constrains microtubule growth differ dramatically for the two scenarios, cf. Fig~\ref{fig:comparison}. For the neutral scenario, this parameter regime is relatively small and, in particular, limited to slow growth rates. It is characterised by relatively tight length control~\cite{Melbinger2012}. In contrast, for the inhibition scenario, the regime where length regulation is possible is extremely broad and includes high growth rates, however, at the cost of accurate length control: microtubule dynamics is intermittent with extended periods of microtubule growth and shrinkage reminiscent of microtubule dynamic instability. Therefore, in view of the regulation of microtubule length, these findings suggest the inhibition scenario as a mechanism for large length fluctuations, while the neutral scenario provides a mechanism for precise length control. To test these theoretical ideas, we suggest experiments which vary the protein concentration of kinesin-8, tubulin, and XMAP215. The specific predictions of our theory will allow to discern between different molecular mechanism at the microtubule tip simply by analysing how changes in the concentrations affect macroscopic quantities like the microtubule length and the speed of microtubule growth and shrinkage. 

Besides its biological relevance, our study also contributes to the field of driven diffusive systems. We not only show, how systems with a dynamic length can be treated analytically, but the technique we propose also gives conceptual insights in the determination of boundary-induced phases. This is achieved by extending the extremal current principle~\cite{Krug1991} to  dynamic systems. For instance, we found that a shock forming dynamically at the right boundary (not in bulk) determines whether the system is in the IN or EX phase. In addition, we could identify an unstable region in the phase diagram (between EX and MC phase for the inhibition scenario), where the system does not only depend on the boundaries, but also on the initial conditions. This behaviour is to our knowledge not common for driven diffusive systems, and an interesting topic for future studies. Even though the main dynamic behaviour, as microtubule length, is governed by currents which are determined by the boundaries, also bulk phenomena as observed in Refs.~\cite{Lipowsky2001, Parmeggiani2003, Leduc2012}, especially for lattice length fluctuations. We restricted our analysis to boundary induced transitions, leaving it as a challenge for the future to capture also the bulk dynamics of motors on the microtubule.

From a broader perspective, the presented findings support the view that length-dependent disassembly and/or assembly rates due to molecular motor transport are likely to constitute a general mechanism to influence the length of one-dimensional structures in biology regardless of mechanistic details~\cite{Marshall2004}. Specifically, microtubule tips are \enquote{crowded} spots in the cell, where space limitations for protein binding, inferring mutual exclusion, are relevant factors. Future experimental work needs to study dwell times of molecules at microtubule tips at the highest possible accuracy, because dwell times encode important information about the underlying molecular interaction networks~\cite{Li2013b}. Similarly it will be important to learn more about interactions of molecular motors with the microtubule~\cite{Vilfan2001,Roos2008} during dynamic instability~\cite{Gardner2011a,Kuan2013,Li2013} and with networks of microtubules~\cite{Neri2013,Greulich2012}.
\section*{Acknowledgments}\vspace{-.3cm}
We thank Matthias Rank for helpful comments on the manuscript. 
This project was supported by the Deutsche Forschungsgemeinschaft in the framework of the SFB~863.


\begin{thebibliography}{61}
\providecommand{\natexlab}[1]{#1}

\bibitem[{Desai \& Mitchison(1997)}]{Desai1997}
Desai, A. \& Mitchison, T. 1997 Microtubule polymerization dynamics.
\newblock \textit{Annu. Rev. Cell Dev. Biol.} \textbf{13} 83--117, (doi:\href{http://dx.doi.org/10.1146/annurev.cellbio.13.1.83}{10.1146/annurev.cellbio.13.1.83}).

\bibitem[{Goshima \& Scholey(2010)}]{Goshima2010}
Goshima, G. \& Scholey, J.~M. 2010 {Control of mitotic spindle length.}
\newblock \textit{Ann. Rev. Cell Dev. Biol.} \textbf{26} 21--57, (doi:\href{http://dx.doi.org/10.1146/annurev-cellbio-100109-104006}{10.1146/annurev-cellbio-100109-104006}).

\bibitem[{de~Forges, Bouissou \& Perez(2012)}]{deForges2012}
de~Forges, H., Bouissou, A. \& Perez, F. 2012 Interplay between microtubule
  dynamics and intracellular organization.
\newblock \textit{Int. J. Biochem. \& Cell Biol.} \textbf{44} 266 -- 274,
  (doi:\href{http://dx.doi.org/10.1016/j.biocel.2011.11.009.}{10.1016/j.biocel.2011.11.009}).

\bibitem[{Tischer, Brunner \& Dogterom(2009)}]{Tischer2009}
Tischer, C., Brunner, D. \& Dogterom, M. 2009 {Force- and kinesin-8-dependent
  effects in the spatial regulation of fission yeast microtubule dynamics.}
\newblock \textit{Mol. Syst. Biol.} \textbf{5} 250, (doi:\href{http://dx.doi.org/10.1038/msb.2009.5}{10.1038/msb.2009.5}).

\bibitem[{Laan, Pavin, Husson, Romet-Lemonne, van Duijn, L\'{o}pez
  et~al.(2012)Laan, Pavin, Husson, Romet-Lemonne, van Duijn, L\'{o}pez, Vale,
  J\"{u}licher, Reck-Peterson \& Dogterom}]{Laan2012}
Laan, L., Pavin, N., Husson, J., Romet-Lemonne, G., van Duijn, M., L\'{o}pez,
  M.~P., Vale, R.~D., J\"{u}licher, F., Reck-Peterson, S.~L. \& Dogterom, M.,
  2012 Cortical dynein controls microtubule dynamics to generate pulling forces
  that position microtubule asters.
\newblock \textit{Cell} \textbf{148} 502--14, (doi:\href{http://dx.doi.org/10.1016/j.cell.2012.01.007}{10.1016/j.cell.2012.01.007}).

\bibitem[{Pavin, Laan, Ma, Dogterom \& J\"{u}licher(2012)}]{Pavin2012}
Pavin, N., Laan, L., Ma, R., Dogterom, M. \& J\"{u}licher, F. 2012
  {Positioning of microtubule organizing centers by cortical pushing and
  pulling forces}.
\newblock \textit{New J. Phys.} \textbf{14} 105025, (doi:\href{http://dx.doi.org/10.1088/1367-2630/14/10/105025}{10.1088/1367-2630/14/10/105025}).

\bibitem[{Howard \& Hyman(2007)}]{Howard2007}
Howard, J. \& Hyman, A.~A. 2007 {Microtubule polymerases and depolymerases.}
\newblock \textit{Curr. Opin. Cell Biol.} \textbf{19} 31--5, (doi:\href{http://dx.doi.org/10.1016/j.ceb.2006.12.009}{10.1016/j.ceb.2006.12.009}).

\bibitem[{Wordeman(2005)}]{Wordeman2005}
Wordeman, L. 2005 {Microtubule-depolymerizing kinesins.}
\newblock \textit{Curr. Opin. Cell Biol.} \textbf{17} 82--8, (doi:\href{http://dx.doi.org/10.1016/j.ceb.2004.12.003}{10.1016/j.ceb.2004.12.003}).

\bibitem[{Akhmanova \& Steinmetz(2008)}]{Akhmanova2008}
Akhmanova, A. \& Steinmetz, M.~O. 2008 Tracking the ends: a dynamic protein
  network controls the fate of microtubule tips.
\newblock \textit{Nat. Rev. Mol. Cell Biol.} \textbf{9} 309--322, (doi:\href{http://dx.doi.org/10.1038/nrm2369}{10.1038/nrm2369}).

\bibitem[{Subramanian \& Kapoor(2012)}]{Subramanian2012}
Subramanian, R. \& Kapoor, T. 2012 Building complexity: Insights into
  self-organized assembly of microtubule-based architectures.
\newblock \textit{Dev. Cell} \textbf{23} 874--885, (doi:\href{http://dx.doi.org/10.1016/j.devcel.2012.10.011}{10.1016/j.devcel.2012.10.011}).

\bibitem[{Varga, Helenius, Tanaka, Hyman, Tanaka \& Howard(2006)}]{Varga2006}
Varga, V., Helenius, J., Tanaka, K., Hyman, A.~A., Tanaka, T.~U. \& Howard, J.,
  2006 {Yeast kinesin-8 depolymerizes microtubules in a length-dependent
  manner.}
\newblock \textit{Nat. Cell Biol.} \textbf{8} 957--62, (doi:\href{http://dx.doi.org/10.1038/ncb1462}{10.1038/ncb1462}).

\bibitem[{Gupta, Carvalho, Roof \& Pellman(2006)}]{Gupta2006}
Gupta, M.~L., Carvalho, P., Roof, D.~M. \& Pellman, D. 2006 {Plus end-specific
  depolymerase activity of Kip3, a kinesin-8 protein, explains its role in
  positioning the yeast mitotic spindle.}
\newblock \textit{Nat. Cell Biol.} \textbf{8} 913--23, (doi:\href{http://dx.doi.org/10.1038/ncb1457}{10.1038/ncb1457}).

\bibitem[{Walczak, Gayek \& Ohi(2013)}]{Walczak2013}
Walczak, C.~E., Gayek, S. \& Ohi, R. 2013 Microtubule-depolymerizing kinesins.
\newblock \textit{Annu. Rev. Cell Dev. Biol.} \textbf{29} 417--441, (doi:\href{http://dx.doi.org/10.1146/annurev-cellbio-101512-122345}{10.1146/annurev-cellbio-101512-122345}).

\bibitem[{Leduc, Padberg-Gehle, Varga, Helbing, Diez \&
  Howard(2012)}]{Leduc2012}
Leduc, C., Padberg-Gehle, K., Varga, V., Helbing, D., Diez, S. \& Howard, J.,
  2012 Molecular crowding creates traffic jams of kinesin motors on
  microtubules.
\newblock \textit{Proc. Natl. Acad. Sci. USA} \textbf{109} 6100--6105, (doi:\href {http://dx.doi.org/10.1073/pnas.1107281109}{10.1073/pnas.1107281109}).

\bibitem[{Cooper, Wagenbach, Asbury \& Wordeman(2010)}]{Cooper2010}
Cooper, J.~R., Wagenbach, M., Asbury, C.~L. \& Wordeman, L. 2010 {Catalysis of
  the microtubule on-rate is the major parameter regulating the depolymerase
  activity of MCAK.}
\newblock \textit{Nat. Struct. Mol. Biol.} \textbf{17} 77--82, (doi:\href{http://dx.doi.org/10.1038/nsmb.1728}{10.1038/nsmb.1728}).

\bibitem[{Varga, Leduc, Bormuth, Diez \& Howard(2009)}]{Varga2009}
Varga, V., Leduc, C., Bormuth, V., Diez, S. \& Howard, J. 2009 {Kinesin-8
  motors act cooperatively to mediate length-dependent microtubule
  depolymerization.}
\newblock \textit{Cell} \textbf{138} 1174--83, (doi:\href{http://dx.doi.org/10.1016/j.cell.2009.07.032}{10.1016/j.cell.2009.07.032}).

\bibitem[{Reese, Melbinger \& Frey(2011)}]{Reese2011}
Reese, L., Melbinger, A. \& Frey, E. 2011 {Crowding of molecular motors
  determines microtubule depolymerization.}
\newblock \textit{Biophys. J.} \textbf{101} 2190--200, (doi:\href{http://dx.doi.org/10.1016/j.bpj.2011.09.009}{10.1016/j.bpj.2011.09.009}).

\bibitem[{Stumpff, Du, English, Maliga, Wagenbach, Asbury et~al.(2011)Stumpff,
  Du, English, Maliga, Wagenbach, Asbury, Wordeman \& Ohi}]{Stumpff2011}
Stumpff, J., Du, Y., English, C., Maliga, Z., Wagenbach, M., Asbury, C.,
  Wordeman, L. \& Ohi, R. 2011 A tethering mechanism controls the processivity
  and kinetochore-microtubule plus-end enrichment of the kinesin-8 {Kif18A}.
\newblock \textit{Mol. Cell} \textbf{43} 764 -- 775, (doi:\href{http://dx.doi.org/10.1016/j.molcel.2011.07.022}{10.1016/j.molcel.2011.07.022}).

\bibitem[{Su, Qiu, Gupta, Pereira-Leal, Reck-Peterson \&
  Pellman(2011)}]{Su2011}
Su, X., Qiu, W., Gupta, M.~L., Pereira-Leal, J.~B., Reck-Peterson, S.~L. \&
  Pellman, D. 2011 {Mechanisms underlying the dual-mode regulation of
  microtubule dynamics by kip3/kinesin-8.}
\newblock \textit{Mol. Cell} \textbf{43} 751--63, (doi:\href{http://dx.doi.org/10.1016/j.molcel.2011.06.027}{10.1016/j.molcel.2011.06.027}).

\bibitem[{Mayr, Storch, Howard \& Mayer(2011)}]{Mayr2011}
Mayr, M.~I., Storch, M., Howard, J. \& Mayer, T.~U. 2011 A non-motor
  microtubule binding site is essential for the high processivity and mitotic
  function of kinesin-8 {Kif18A}.
\newblock \textit{PLoS ONE} \textbf{6} e27471, (doi:\href{http://dx.doi.org/10.1371/journal.pone.0027471}{10.1371/journal.pone.0027471}).

\bibitem[{Weaver, Ems-McClung, Stout, Leblanc, Shaw, Gardner
  et~al.(2011)Weaver, Ems-McClung, Stout, Leblanc, Shaw, Gardner \&
  Walczak}]{Weaver2011}
Weaver, L.~N., Ems-McClung, S.~C., Stout, J.~R., Leblanc, C., Shaw, S.~L.,
  Gardner, M.~K. \& Walczak, C.~E. 2011 {{Kif18A} Uses a Microtubule Binding
  Site in the Tail for Plus-End Localization and Spindle Length Regulation.}
\newblock \textit{Curr. Biol.} \textbf{21} 1500--6, (doi:\href{http://dx.doi.org/10.1016/j.cub.2011.08.005}{10.1016/j.cub.2011.08.005}).

\bibitem[{Su, Ohi \& Pellman(2012)}]{Su2012}
Su, X., Ohi, R. \& Pellman, D. 2012 Move in for the kill: motile microtubule
  regulators.
\newblock \textit{Trends Cell Biol.} \textbf{22} 567--575, (doi:\href{http://dx.doi.org/10.1016/j.tcb.2012.08.003}{10.1016/j.tcb.2012.08.003}).

\bibitem[{Gard \& Kirschner(1987)}]{Gard1987}
Gard, D.~L. \& Kirschner, M.~W. 1987 {A microtubule-associated protein from
  Xenopus eggs that specifically promotes assembly at the plus-end.}
\newblock \textit{J. Cell Biol.} \textbf{105} 2203--15, (doi:\href{http://dx.doi.org/10.1083/jcb.105.5.2203}{10.1083/jcb.105.5.2203}).

\bibitem[{Vasquez, Gard \& Cassimeris(1994)}]{Vasquez1994}
Vasquez, R.~J., Gard, D.~L. \& Cassimeris, L. 1994 Xmap from xenopus eggs
  promotes rapid plus end assembly of microtubules and rapid microtubule
  polymer turnover.
\newblock \textit{J. Cell Biol.} \textbf{127} 985--993, (doi:\href{http://dx.doi.org/10.1083/jcb.127.4.985}{10.1083/jcb.127.4.985}).

\bibitem[{Tournebize, Popov, Kinoshita, Ashford, Rybina, Pozniakovsky
  et~al.(2000)Tournebize, Popov, Kinoshita, Ashford, Rybina, Pozniakovsky,
  Mayer, Walczak, Karsenti \& Hyman}]{Tournebize2000}
Tournebize, R., Popov, A., Kinoshita, K., Ashford, A.~J., Rybina, S.,
  Pozniakovsky, A., Mayer, T.~U., Walczak, C.~E., Karsenti, E. \& Hyman, A.~A.,
  2000 {Control of microtubule dynamics by the antagonistic activities of
  {XMAP215} and XKCM1 in Xenopus egg extracts.}
\newblock \textit{Nat. Cell Biol.} \textbf{2} 13--9, (doi:\href{http://dx.doi.org/10.1038/71330}{10.1038/71330}).

\bibitem[{Kinoshita, Habermann \& Hyman(2002)}]{Kinoshita2002}
Kinoshita, K., Habermann, B. \& Hyman, A.~A. 2002 {{XMAP215}: a key component
  of the dynamic microtubule cytoskeleton}.
\newblock \textit{Trends Cell Biol.} \textbf{12} 267--273, (doi:\href{http://dx.doi.org/10.1016/S0962-8924(02)02295-X}{10.1016/S0962-8924(02)02295-X}).

\bibitem[{Brouhard, Stear, Noetzel, Al-Bassam, Kinoshita, Harrison
  et~al.(2008)Brouhard, Stear, Noetzel, Al-Bassam, Kinoshita, Harrison, Howard
  \& Hyman}]{Brouhard2008}
Brouhard, G.~J., Stear, J.~H., Noetzel, T.~L., Al-Bassam, J., Kinoshita, K.,
  Harrison, S.~C., Howard, J. \& Hyman, A.~A. 2008 {{XMAP215} is a processive
  microtubule polymerase.}
\newblock \textit{Cell} \textbf{132} 79--88, (doi:\href{http://dx.doi.org/10.1016/j.cell.2007.11.043}{10.1016/j.cell.2007.11.043}).

\bibitem[{Li, Moriwaki, Tani, Watanabe, Kaibuchi \& Goshima(2012)}]{Li2012}
Li, W., Moriwaki, T., Tani, T., Watanabe, T., Kaibuchi, K. \& Goshima, G. 2012
  Reconstitution of dynamic microtubules with drosophila {XMAP215}, {EB1}, and
  sentin.
\newblock \textit{J. Cell Biol.} \textbf{199} 849--862, (doi:\href{http://dx.doi.org/10.1083/jcb.201206101}{10.1083/jcb.201206101}).

\bibitem[{Zanic, Widlund, Hyman \& Howard(2013)}]{Zanic2013}
Zanic, M., Widlund, P.~O., Hyman, A.~A. \& Howard, J. 2013 {Synergy between
  {XMAP215} and {EB1} increases microtubule growth rates to physiological
  levels.}
\newblock \textit{Nat. Cell Biol.} \textbf{15} 688--93, (doi:\href{http://dx.doi.org/10.1038/ncb2744}{10.1038/ncb2744}).

\bibitem[{Komarova, {De Groot}, Grigoriev, Gouveia, Munteanu, Schober
  et~al.(2009)Komarova, {De Groot}, Grigoriev, Gouveia, Munteanu, Schober,
  Honnappa, Buey, Hoogenraad, Dogterom et~al.}]{Komarova2009}
Komarova, Y., {De Groot}, C., Grigoriev, I., Gouveia, S., Munteanu, E.,
  Schober, J., Honnappa, S., Buey, R., Hoogenraad, C., Dogterom, M. et~al.,
  2009 {Mammalian end binding proteins control persistent microtubule growth}.
\newblock \textit{J. Cell Biol.} \textbf{184} 691--706, (doi:\href{http://dx.doi.org/10.1083/jcb.200807179}{10.1083/jcb.200807179}).

\bibitem[{Kinoshita, Arnal, Desai, Drechsel \& Hyman(2001)}]{Kinoshita2001}
Kinoshita, K., Arnal, I., Desai, A., Drechsel, D.~N. \& Hyman, A.~A. 2001
  {Reconstitution of physiological microtubule dynamics using purified
  components.}
\newblock \textit{Science} \textbf{294} 1340--3, (doi:\href{http://dx.doi.org/10.1126/science.1064629}{10.1126/science.1064629}).

\bibitem[{Niwa, Nakajima, Miki, Minato, Wang \& Hirokawa(2012)}]{Niwa2012}
Niwa, S., Nakajima, K., Miki, H., Minato, Y., Wang, D. \& Hirokawa, N. 2012
  Kif19a is a microtubule-depolymerizing kinesin for ciliary length control.
\newblock \textit{Dev. Cell} \textbf{23} 1167--1175, (doi:\href{http://dx.doi.org/10.1016/j.devcel.2012.10.016}{10.1016/j.devcel.2012.10.016}).

\bibitem[{Melbinger, Reese \& Frey(2012)}]{Melbinger2012}
Melbinger, A., Reese, L. \& Frey, E. 2012 Microtubule length regulation by
  molecular motors.
\newblock \textit{Phys. Rev. Lett.} \textbf{108} 258104, (doi:\href{http://dx.doi.org/10.1103/PhysRevLett.108.258104}{10.1103/PhysRevLett.108.258104}).

\bibitem[{Krapivsky, Redner \& Ben-Naim(2010)}]{Krapivsky2010}
Krapivsky, P.~L., Redner, S. \& Ben-Naim, E. 2010 \textit{A kinetic view of statistical physics}.
\newblock Cambridge University Press.

\bibitem[{Chou, Mallick \& Zia(2011)}]{Chou2011}
Chou, T., Mallick, K. \& Zia, R. K.~P. 2011 Non-equilibrium statistical
  mechanics: from a paradigmatic model to biological transport.
\newblock \textit{Rep. Prog. Phys.} \textbf{74} 116601, (doi:\href{http://dx.doi.org/10.1088/0034-4885/74/11/116601}{10.1088/0034-4885/74/11/116601}).

\bibitem[{Lipowsky, Klumpp \& Nieuwenhuizen(2001)}]{Lipowsky2001}
Lipowsky, R., Klumpp, S. \& Nieuwenhuizen, T. 2001 {Random Walks of
  Cytoskeletal Motors in Open and Closed Compartments}.
\newblock \textit{Phys. Rev. Lett.} \textbf{87} 108101, (doi:\href{http://dx.doi.org/10.1103/PhysRevLett.87.108101}{10.1103/PhysRevLett.87.108101}).

\bibitem[{Parmeggiani, Franosch \& Frey(2003)}]{Parmeggiani2003}
Parmeggiani, A., Franosch, T. \& Frey, E. 2003 {Phase Coexistence in Driven
  One-Dimensional Transport}.
\newblock \textit{Phys. Rev. Lett.} \textbf{90} 86601, (doi:\href{http://dx.doi.org/10.1103/PhysRevLett.90.086601}{10.1103/PhysRevLett.90.086601}).

\bibitem[{Parmeggiani, Franosch \& Frey(2004)}]{Parmeggiani2004}
Parmeggiani, A., Franosch, T. \& Frey, E. 2004 {Totally asymmetric simple
  exclusion process with Langmuir kinetics}.
\newblock \textit{Phys. Rev. E} \textbf{70} 46101, (doi:\href{http://dx.doi.org/10.1103/PhysRevE.70.046101}{10.1103/PhysRevE.70.046101}).

\bibitem[{Gillespie(2007)}]{Gillespie2007}
Gillespie, D.~T. 2007 Stochastic simulation of chemical kinetics.
\newblock \textit{Annu. Rev. Phys. Chem.} \textbf{58} 35--55, (doi:\href{http://dx.doi.org/10.1146/annurev.physchem.58.032806.104637}{10.1146/annurev.physchem.58.032806.104637}).

\bibitem[{Mitchison \& Kirschner(1984)}]{Mitchison1984}
Mitchison, T. \& Kirschner, M. 1984 {Dynamic instability of microtubule
  growth.}
\newblock \textit{Nature} \textbf{312} 237--42, (doi:\href{http://dx.doi.org/10.1038/312237a0}{10.1038/312237a0}).

\bibitem[{Derrida, Domany \& Mukamel(1992)}]{Derrida1992}
Derrida, B., Domany, E. \& Mukamel, D. 1992 An exact solution of a
  one-dimensional asymmetric exclusion model with open boundaries.
\newblock \textit{J. Stat. Phys.} \textbf{69} 667--687, (doi:\href{http://dx.doi.org/10.1007/BF01050430}{10.1007/BF01050430}).

\bibitem[{Sch\"{u}tz \& Domany(1993)}]{Schuetz1993}
Sch\"{u}tz, G. \& Domany, E. 1993 Phase transitions in an exactly soluble
  one-dimensional exclusion process.
\newblock \textit{J. Stat. Phys.} \textbf{72} 277--296, (doi:\href{http://dx.doi.org/10.1007/BF01048050}{10.1007/BF01048050}).

\bibitem[{Hager, Krug, Popkov \& Sch\"utz(2001)}]{Hager2001}
Hager, J.~S., Krug, J., Popkov, V. \& Sch\"utz, G.~M. 2001 Minimal current
  phase and universal boundary layers in driven diffusive systems.
\newblock \textit{Phys. Rev. E} \textbf{63} 056110, (doi:\href{http://dx.doi.org/10.1103/PhysRevE.63.056110}{10.1103/PhysRevE.63.056110}).

\bibitem[{Krug(1991)}]{Krug1991}
Krug, J. 1991 {Boundary-induced phase transitions in driven diffusive
  systems}.
\newblock \textit{Phys. Rev. Lett.} \textbf{67} 1882--1885, (doi:\href{http://dx.doi.org/10.1103/PhysRevLett.67.1882}{10.1103/PhysRevLett.67.1882}).

\bibitem[{Kolomeisky, Sch\"{u}tz, Kolomeisky \& Straley(1998)}]{Kolomeisky1998}
Kolomeisky, A.~B., Sch\"{u}tz, G.~M., Kolomeisky, E.~B. \& Straley, J.~P. 1998
  Phase diagram of one-dimensional driven lattice gases with open boundaries.
\newblock \textit{J. Phys. A: Math. Gen.} \textbf{31} 6911, (doi:\href{http://dx.doi.org/10.1088/0305-4470/31/33/003}{10.1088/0305-4470/31/33/003}).

\bibitem[{Popkov \& Sch\"{u}tz(1999)}]{Popkov1999}
Popkov, V. \& Sch\"{u}tz, G.~M. 1999 {Steady-state selection in driven
  diffusive systems with open boundaries}.
\newblock \textit{Europhys. Lett.} \textbf{48} 257--263, (doi:\href{http://dx.doi.org/10.1209/epl/i1999-00474-0}{10.1209/epl/i1999-00474-0}).

\bibitem[{Wood(2009)}]{Wood2009}
Wood, A.~J. 2009 A totally asymmetric exclusion process with stochastically
  mediated entrance and exit.
\newblock \textit{J. Phys. A: Math. Theor.} \textbf{42} 445002, (doi:\href{http://dx.doi.org/10.1088/1751-8113/42/44/445002}{10.1088/1751-8113/42/44/445002}).

\bibitem[{Dogterom \& Leibler(1993)}]{Dogterom1993}
Dogterom, M. \& Leibler, S. 1993 {Physical aspects of the growth and
  regulation of microtubule structures}.
\newblock \textit{Phys. Rev. Lett.} \textbf{70} 1347--1350, (doi:\href{http://dx.doi.org/10.1103/PhysRevLett.70.1347}{10.1103/PhysRevLett.70.1347}).

\bibitem[{Marshall(2004)}]{Marshall2004}
Marshall, W.~F. 2004 {Cellular length control systems.}
\newblock \textit{Ann. Rev. Cell Dev. Biol.} \textbf{20} 677--93, (doi:\href{http://dx.doi.org/10.1146/annurev.cellbio.20.012103.094437}{10.1146/annurev.cellbio.20.012103.094437}).

\bibitem[{Li \& Kolomeisky(2013{\natexlab{a}})}]{Li2013b}
Li, X. \& Kolomeisky, A.~B. 2013{\natexlab{a}} Mechanisms and topology
  determination of complex chemical and biological network systems from
  first-passage theoretical approach.
\newblock \textit{J. Chem. Phys.} \textbf{139} 144106, (doi:\href{http://dx.doi.org/10.1063/1.4824392}{10.1063/1.4824392}).

\bibitem[{Vilfan, Frey, Schwabl, Thorm\"{a}hlen, Song \&
  Mandelkow(2001)}]{Vilfan2001}
Vilfan, A., Frey, E., Schwabl, F., Thorm\"{a}hlen, M., Song, Y.~H. \&
  Mandelkow, E. 2001 {Dynamics and cooperativity of microtubule decoration by
  the motor protein kinesin.}
\newblock \textit{J. Mol. Biol.} \textbf{312} 1011--26, (doi:\href{http://dx.doi.org/10.1006/jmbi.2001.5020}{10.1006/jmbi.2001.5020}).

\bibitem[{Roos, Camp\`{a}s, Montel, Woehlke, Spatz, Bassereau et~al.(2008)Roos,
  Camp\`{a}s, Montel, Woehlke, Spatz, Bassereau \& Cappello}]{Roos2008}
Roos, W., Camp\`{a}s, O., Montel, F., Woehlke, G., Spatz, J., Bassereau, P. \&
  Cappello, G. 2008 {Dynamic kinesin-1 clustering on microtubules due to
  mutually attractive interactions}.
\newblock \textit{Phys. Biol.} \textbf{5} 46004, (doi:\href{http://dx.doi.org/10.1088/1478-3975/5/4/046004}{10.1088/1478-3975/5/4/046004}).

\bibitem[{Gardner, Zanic, Gell, Bormuth \& Howard(2011)}]{Gardner2011a}
Gardner, M.~K., Zanic, M., Gell, C., Bormuth, V. \& Howard, J. 2011
  {Depolymerizing Kinesins Kip3 and MCAK Shape Cellular Microtubule
  Architecture by Differential Control of Catastrophe}.
\newblock \textit{Cell} \textbf{147} 1092--1103, (doi:\href{http://dx.doi.org/10.1016/j.cell.2011.10.037}{10.1016/j.cell.2011.10.037}).

\bibitem[{Kuan \& Betterton(2013)}]{Kuan2013}
Kuan, H.-S. \& Betterton, M.~D. 2013 Biophysics of filament length regulation
  by molecular motors.
\newblock \textit{Phys. Biol.} \textbf{10} 036004, (doi:\href{http://dx.doi.org/10.1088/1478-3975/10/3/036004}{10.1088/1478-3975/10/3/036004}).

\bibitem[{Li \& Kolomeisky(2013{\natexlab{b}})}]{Li2013}
Li, X. \& Kolomeisky, A.~B. 2013{\natexlab{b}} Theoretical analysis of
  microtubules dynamics using a physical-chemical description of hydrolysis.
\newblock \textit{J. Phys. Chem. B} \textbf{117} 9217--9223, (doi:\href{http://dx.doi.org/10.1021/jp404794f}{10.1021/jp404794f}).

\bibitem[{Neri, Kern \& Parmeggiani(2013)}]{Neri2013}
Neri, I., Kern, N. \& Parmeggiani, A. 2013 Modeling cytoskeletal traffic: An
  interplay between passive diffusion and active transport.
\newblock \textit{Phys. Rev. Lett.} \textbf{110} 098102, (doi:\href{http://dx.doi.org/10.1103/PhysRevLett.110.098102}{10.1103/PhysRevLett.110.098102}).

\bibitem[{Greulich, Ciandrini, Allen \& Romano(2012)}]{Greulich2012}
Greulich, P., Ciandrini, L., Allen, R.~J. \& Romano, M.~C. 2012 Mixed
  population of competing totally asymmetric simple exclusion processes with a
  shared reservoir of particles.
\newblock \textit{Phys. Rev. E} \textbf{85} 011142, (doi:\href{http://dx.doi.org/10.1103/PhysRevE.85.011142}{10.1103/PhysRevE.85.011142}).

\bibitem[{Pierobon, Mobilia, Kouyos \& Frey(2006)}]{Pierobon2006}
Pierobon, P., Mobilia, M., Kouyos, R. \& Frey, E. 2006 {Bottleneck-induced
  transitions in a minimal model for intracellular transport}.
\newblock \textit{Phys. Rev. E} \textbf{74} 31906, (doi:\href{http://dx.doi.org/10.1103/PhysRevE.74.031906}{10.1103/PhysRevE.74.031906}).

\bibitem[{Stumpff, von Dassow, Wagenbach, Asbury \&
  Wordeman(2008)}]{Stumpff2008}
Stumpff, J., von Dassow, G., Wagenbach, M., Asbury, C. \& Wordeman, L. 2008
  {The kinesin-8 motor {Kif18A} suppresses kinetochore movements to control
  mitotic chromosome alignment.}
\newblock \textit{Dev. Cell} \textbf{14} 252--62, (doi:\href{http://dx.doi.org/10.1016/j.devcel.2007.11.014}{10.1016/j.devcel.2007.11.014}).

\bibitem[{Gardner, Charlebois, Janosi, Howard, Hunt \&
  Odde(2011)}]{Gardner2011}
Gardner, M.~K., Charlebois, B.~D., Janosi, I.~M., Howard, J., Hunt, A.~J. \&
  Odde, D.~J. 2011 {Rapid Microtubule Self-Assembly Kinetics}.
\newblock \textit{Cell} \textbf{146} 582--592, (doi:\href{http://dx.doi.org/10.1016/j.cell.2011.06.053}{10.1016/j.cell.2011.06.053}).

\bibitem[{Du, English \& Ohi(2010)}]{Du2010}
Du, Y., English, C.~A. \& Ohi, R. 2010 {The kinesin-8 {Kif18A} dampens
  microtubule plus-end dynamics.}
\newblock \textit{Curr. Biol.} \textbf{20} 374--80, (doi:\href{http://dx.doi.org/10.1016/j.cub.2009.12.049}{10.1016/j.cub.2009.12.049}).

\end{thebibliography}

\end{document}